%% file: main.tex
\documentclass[preprint,12pt]{elsarticle}

\usepackage{subfiles}
\usepackage{array}
\usepackage{colortbl}
\usepackage{booktabs}
\usepackage{graphicx}
\usepackage{multirow}
\usepackage{amsmath,amssymb,amsfonts}
\usepackage{amsthm}
\usepackage{xcolor}
\usepackage{textcomp}
\usepackage{algorithm}
\usepackage{algpseudocode}
\usepackage{listings}
\usepackage{bm}
\usepackage{caption}
\usepackage{paralist}
\usepackage{ragged2e}
\usepackage{rotating}
\usepackage{adjustbox}
\usepackage{pdflscape}
\usepackage{makecell}
\usepackage{url}
\usepackage{hyperref}
\hypersetup{unicode,breaklinks=true}

\journal{Information Systems}

\lstdefinestyle{customsql}{
  language=SQL,
  basicstyle=\ttfamily\small,
  breaklines=true,
  frame=single,
  columns=flexible
}
\lstdefinestyle{customjson}{
  language=Java,
  basicstyle=\ttfamily\small,
  breaklines=true,
  frame=single,
  columns=flexible
}

\input{macros}

\begin{document}

\subfile{abstract}

\include{chapters/chap01}
\include{chapters/chap02}
\include{chapters/chap03}
\include{chapters/chap04}
\include{chapters/chap05}
\include{chapters/chap06}
\include{chapters/epilog}

\section*{Acknowledgment}
Supported by the GAČR grant no. 23-07781S. 

\bibliographystyle{elsarticle-num}
\bibliography{bibliography}

\end{document}

%% file: macros.tex
\newcommand{\old}[1]{}
\newcommand{\new}[1]{#1}

%% file: abstract.tex
\begin{frontmatter}

\title{Benchmark Engineering as a Design Instrument for Heterogeneous Information Systems}

\author[inst1]{Jáchym Bártík\corref{cor1}}
\ead{jachym.bartik@matfyz.cuni.cz}

\author[inst1]{Alžběta Šrůtková}
\ead{srutkova.alzbeta@gmail.com}

\author[inst1]{Irena Holubová}
\ead{irena.holubova@matfyz.cuni.cz}

\cortext[cor1]{Corresponding author.}

\address[inst1]{Department of Software Engineering, Charles University, Malostranské nám.\ 25, \\ 
Praha 118 00, Czech Republic}

\begin{abstract}
\input{abstract-content.tex}
\end{abstract}

\begin{keyword}
multi-model data management \sep
benchmark engineering \sep
schema evolution \sep
heterogeneous information systems \sep
cross-model transformation \sep
reproducible benchmarking
\end{keyword}

\end{frontmatter}

%% file: abstract-content.tex
Contemporary information systems operate in heterogeneous and continuously evolving data environments, where representation choices and structural redesign decisions strongly influence system behavior. Existing benchmarking approaches, however, rely mostly on static datasets and fixed schemas, providing limited support for analyzing architectural trade-offs or guiding evolution in multi-model settings.

This paper introduces \emph{TransforMMer}, a framework for evolution-aware and representation-aware benchmark engineering in heterogeneous information systems. The approach treats benchmark construction as a systematic design process: starting from raw data, inferring structure, refining it conceptually, and generating comparable dataset variants across relational, document, and graph systems. The framework is grounded in a unified representation that enables explicit modeling of schemas and cross-model mappings and supports reproducible transformations across alternative representations.

We position benchmarking as a system-design tool for evaluating architectural and representation-level decisions in evolving information systems, rather than as a static comparison of database engines. Through controlled benchmark construction scenarios on real-world datasets, we demonstrate how structural redesign steps---such as embedding, enrichment, and hybrid partitioning---affect observed query costs across systems. The results show that performance differences emerge primarily from the interaction between workload and representation design.

By enabling systematic generation of structurally distinct yet semantically \old{equivalent}\new{aligned} dataset variants, the proposed approach connects conceptual data modeling with empirical system evaluation and supports reproducible, evolution-aware analysis of heterogeneous information systems.

%% file: chapters/chap01.tex
\section{Introduction and Motivation}
\label{sec:intro}

The emergence of Big Data has significantly changed the landscape of modern information systems. Contemporary applications increasingly operate over heterogeneous data environments that combine relational, document, key-value, and graph data models. This diversity has led to the development of multi-model database systems (MMDBs), which integrate different data abstractions within a single platform and support a broad spectrum of workloads.

While multi-model systems improve flexibility and enable model-specific optimizations, they also introduce substantial complexity. Differences in schema structures, storage strategies, indexing mechanisms, and query languages complicate both system design and evaluation. This challenge becomes even more pronounced in environments where schemas evolve over time, data representations change, and systems integrate multiple storage technologies. As a result, systematic evaluation of heterogeneous information systems remains difficult.

Benchmarking has traditionally served as a primary method for evaluating database systems and data processing platforms. Established benchmarks, such as TPC~\cite{TPC} for relational systems, LDBC~\cite{LDBC} for graph databases, and YCSB~\cite{YCSB} for key-value stores, enable controlled, reproducible experiments. However, these benchmarks typically assume fixed schemas, stable workloads, and isolated data models. Consequently, they do not sufficiently capture the dynamic and heterogeneous nature of contemporary information systems. Even recent multi-model benchmarks remain largely static, domain-specific, and difficult to adapt to evolving datasets or alternative structural representations.

In real-world information systems, data and schemas rarely remain static.
\old{Structural redesign is common as applications evolve: attributes are added or removed, datasets are enriched from additional sources, and storage layouts are reorganized across technologies.}\new{Two major sources of dynamics shape such systems: conceptual schema changes---such as adding or removing attributes, enriching datasets from additional sources, or reorganizing storage layouts---and changes in application workloads and access patterns, such as new query types, shifted analytical demands, or the integration of new data consumers. Both kinds of change frequently require restructuring the system's logical representations. Notably, workload-driven adaptations need not involve any modification to the underlying conceptual schema; alternative logical mappings are sufficient to accommodate new access patterns.} Such changes often influence performance characteristics and system behavior, yet they are rarely captured within benchmarking methodologies. Evaluation is therefore frequently conducted in an ad-hoc manner, and results are difficult to reproduce when structural conditions change.

We argue that benchmarking in heterogeneous information systems should be treated as a systematic engineering process rather than as a fixed dataset specification. Such a process should support controlled schema and data evolution, \new{workload-driven adaptation of logical representations,} alternative structural representations of the same conceptual data, and reproducible transformation workflows across multiple database systems. Enabling these capabilities allows benchmarking to reflect realistic system evolution and provides deeper insight into trade-offs between data models, storage strategies, and workloads.

To address these challenges, this paper introduces an evolution-aware approach to multi-model benchmarking implemented in the \emph{TransforMMer} framework. The framework integrates schema inference, unified conceptual representation, cross-model transformation, and versioned dataset generation within a single workflow. Instead of providing a single static benchmark dataset, the proposed approach enables systematic generation of multiple structurally distinct yet semantically \old{equivalent}\new{aligned} dataset variants that can be evaluated across heterogeneous database systems.

Unlike traditional benchmarks that compare database technologies under fixed schemas, our approach enables exploration of how structural transformations and representation choices influence observed system behavior. This perspective reflects realistic information systems practice, where schema evolution, data integration, and performance-driven redesign are continuous processes rather than one-time decisions.

\paragraph*{Research Questions}

To guide the design and evaluation of the proposed framework, this work addresses the following research questions:

\begin{itemize}
    \item \textbf{RQ1:} How can benchmark datasets for heterogeneous information systems be systematically generated from real-world data while supporting multiple data models and target platforms?

    \item \textbf{RQ2:} How do alternative structural representations of the same conceptual dataset influence observed performance characteristics across relational, document, and graph database systems?

    \item \textbf{RQ3:} To what extent can evolution-aware benchmark generation reveal system-level trade-offs that remain hidden in traditional static benchmark settings?

    \item \textbf{RQ4:} How can transformation-driven benchmark construction support reproducible evaluation and iterative redesign within the information systems life cycle?
\end{itemize}

While benchmarking is traditionally associated with database system evaluation, its role in information systems engineering is broader. In modern data-intensive information systems, structural design decisions, such as schema restructuring, data integration, and model selection, directly influence system performance, maintainability, and the cost of evolution. 
We therefore position evolution-aware benchmarking as a design-support instrument for heterogeneous information systems rather than merely a performance comparison tool. By enabling controlled exploration of alternative structural representations and storage strategies, the proposed framework supports architectural decision-making 
throughout the information systems life cycle.

\paragraph*{Contributions}

This paper makes the following contributions:

\begin{itemize}
    \item We conceptualize evolution-aware benchmarking as an architectural design and evaluation methodology for heterogeneous information systems.
    \item We propose a unified framework that integrates schema inference, conceptual modeling, cross-model transformation, and versioned dataset generation into a reproducible workflow.
    \item We demonstrate how controlled structural transformations enable systematic exploration of performance trade-offs across relational, document, and graph database systems.
    \item We introduce \emph{DaRe}, a repository of datasets, mappings, and transformation artifacts that supports transparency, reuse, and reproducible experimentation.
\end{itemize}

This work contributes to Information Systems research by positioning benchmarking as a design-support methodology for architectural decision-making in heterogeneous data-intensive systems.

\paragraph*{Outline}
The remainder of this paper is structured as follows. Section~\ref{sec:related} reviews related work in data sources, generators, and benchmarking. Section~\ref{sec:theory} introduces the conceptual foundations of the approach. Section~\ref{sec:tool} presents the design and architecture of the framework. Section~\ref{sec:experiments} describes the experimental evaluation. Section~\ref{sec:discussion} discusses implications for information systems research and practice. Section~\ref{sec:concl} concludes the paper and outlines directions for future work.

%% file: chapters/chap02.tex
\section{Related Work and Positioning}
\label{sec:related}

Benchmarking and synthetic data generation are central to evaluating data management technologies and information systems. Existing approaches span traditional single-model benchmarks, synthetic data generators, and more recent multi-model benchmarking initiatives. However, most existing solutions assume static schemas, predefined workloads, and fixed data representations. This section reviews existing approaches and positions the proposed framework within this landscape.

\subsection{Traditional Benchmarks and Data Generators}

Benchmarking has long been used to evaluate the performance and scalability of database systems. Established benchmarks such as TPC-H and TPC-DS provide standardized relational workloads and large-scale data generators for analytical processing. Similarly, YCSB focuses on key-value and cloud-serving systems, while LDBC provides graph workloads for social network scenarios. XML-oriented benchmarks such as XMark~\cite{schmidt2002xmark} and RDF benchmarks such as LUBM~\cite{guo2005lubm} support evaluation of semi-structured and semantic data systems.

These benchmarks offer well-defined schemas and reproducible workloads, which make them valuable for controlled experiments. However, they typically assume static schemas and a single dominant data model. As a result, they provide limited support for evaluating heterogeneous information systems in which multiple data models coexist and evolve over time.

Synthetic data generators provide additional flexibility by enabling the creation of datasets with configurable size and structure. Tools such as SDV~\cite{SDV}, Faker~\cite{Faker}, and domain-specific generators support the generation of relational, document, or graph data. Graph-oriented generators such as gMark~\cite{bagan2016gmark} enable configurable graph structures and workloads. Despite their flexibility, most generators focus on individual data models and do not provide unified support for heterogeneous multi-model environments.

\subsection{Multi-Model Benchmarks}

The increasing adoption of heterogeneous data management platforms has motivated the development of multi-model benchmarks. BigBench~\cite{ghazal2013bigbench} integrates structured, semi-structured, and unstructured data within a unified benchmark scenario. UniBench~\cite{Zhang2019} and M2Bench~\cite{Kim2022} extend this direction by supporting multiple data models, including relational, graph, document, and key-value representations. MMSBench~\cite{lengweiler2023mmsbench} focuses on network monitoring scenarios and heterogeneous data ingestion.

These approaches represent an important progress toward heterogeneous benchmarking. Nevertheless, most existing multi-model benchmarks remain tied to predefined schemas and specific application domains. They typically provide limited support for systematic schema evolution, controlled structural transformations, or the generation of multiple representation variants of the same dataset. Consequently, they are less suitable for studying how structural redesign and representation choices influence system behavior over time.

\subsection{Research Gap}

Contemporary information systems rarely operate under fixed schemas or static data representations. Instead, they evolve continuously in response to changing workloads, integration requirements, and performance constraints. Structural redesign steps such as embedding, normalization changes, enrichment from additional data sources, or distribution across multiple storage technologies are common in practice. However, existing benchmarking approaches offer limited support for systematically evaluating such evolution.

In particular, three limitations can be identified in current benchmarking and data generation approaches:

\begin{itemize}
    \item \textbf{Limited support for evolution.} Most benchmarks assume static schemas and do not support controlled generation of dataset versions reflecting structural change.
    \item \textbf{Lack of representation variability.} Existing benchmarks typically provide a single structural representation of the data, limiting exploration of alternative designs.
    \item \textbf{Insufficient reproducibility of transformations.} Structural modifications applied during system redesign are often ad hoc and difficult to reproduce in benchmarking scenarios.
    \item \new{\textbf{Limited support for workload-driven adaptation.} Existing approaches do not facilitate systematic exploration of alternative logical representations driven by changing access patterns or query workloads.}
\end{itemize}

These limitations highlight the need for benchmarking approaches that treat dataset construction as a configurable, reproducible process rather than a fixed specification.

\subsection{Positioning of the Proposed Approach}

The framework presented in this paper addresses the above limitations by enabling transformation-driven generation of benchmark datasets across heterogeneous data models. Instead of focusing on predefined schemas or domain-specific workloads, the proposed approach supports systematic construction of multiple structural representations of the same conceptual dataset and controlled propagation of structural evolution.

Table~\ref{tab:related-comparison} summarizes the key differences between existing approaches and the proposed framework.

\begin{table}[h]
\caption{Comparison of existing benchmarking and data generation approaches with the proposed framework}
\label{tab:related-comparison}
\centering
\small
\begin{tabular}{p{3cm}p{1.6cm}p{2cm}p{2.4cm}p{2.7cm}}
\toprule
\textbf{Approach} & \textbf{Multi-model support} & \textbf{Schema evolution support} & \textbf{Cross-model transformations} & \textbf{General-purpose} \\
\midrule
TPC-H/TPC-DS & No & No & No & Yes (relational) \\
XMark~\cite{schmidt2002xmark} & No & No & No & XML-specific \\
LUBM~\cite{guo2005lubm} & No & Limited & No & RDF-specific \\
BigBench~\cite{ghazal2013bigbench} & Partial & No & Limited & Domain-specific \\
UniBench~\cite{Zhang2019} & Yes & No & Limited & Domain-specific \\
M2Bench~\cite{Kim2022} & Yes & Limited & Limited & Domain-specific \\
MMSBench~\cite{lengweiler2023mmsbench} & Yes & Limited & Limited & Domain-specific \\
Data generators & Partial & No & No & Model-specific \\
Graph generators & No & Limited & No & Model-specific \\
\midrule
\textbf{Proposed framework} & \textbf{Yes} & \textbf{Yes} & \textbf{Yes} & \textbf{Yes} \\
\bottomrule
\end{tabular}
\end{table}

Unlike existing approaches, the proposed framework treats benchmark construction as a transformation-driven process that supports multiple data models, explicit schema evolution, and reproducible generation of structurally distinct dataset variants. This positioning aligns benchmarking with the realities of contemporary heterogeneous information systems, where structural redesign and incremental evolution are integral parts of system development and operation.

%% file: chapters/chap03.tex
\section{Conceptual Foundations}
\label{sec:theory}

A central challenge in multi-model data management is the consistent representation and transformation of heterogeneous schemas. Modern database systems support relational tables, nested documents, graphs, and key-value structures, yet structural interoperability across these models remains limited. Our benchmarking approach addresses this limitation by introducing a unified structural representation that enables controlled transformations across models and supports reproducible evaluation.

The proposed framework builds on category-theoretic principles and introduces three core constructs:
(i) the \emph{schema category},
(ii) the \emph{instance category}, and
(iii) explicit \emph{mappings} between system-specific representations and the unified structure.
A detailed formal treatment is provided in~\cite{Koupil2022}; here we summarize only the concepts necessary for understanding benchmark construction.

\subsection{Unified Structural Representation}

To reason consistently across heterogeneous models, we introduce model-agnostic terminology:

\begin{itemize}
    \item \old{\textbf{Object}}\new{\textbf{Entity}} -- a stored entity (e.g., tuple, document, node),
    \item \textbf{Property} -- an attribute of an \old{object}\new{entity},
    \item \textbf{Reference} -- a relationship between \old{objects}\new{entities}.
\end{itemize}

These abstractions allow structural reasoning independent of implementation syntax.

Category theory provides the formal basis for this unification. A \emph{category} consists of \emph{objects} and \emph{morphisms} between them, together with composition and identity operations. In our context, objects correspond to structural elements (e.g., entity types), while morphisms represent relationships or dependencies. This abstraction allows heterogeneous schemas to be represented within a single canonical structure.

\paragraph{Running Example}

Consider a simple dataset describing \textit{Users} and \textit{Friends}. Each user also has an \textit{Address}. In a relational database system, this information can be represented using three tables:
\begin{itemize}
    \item \texttt{user(id, name)}
    \item \texttt{friend(user\_id, friend\_id, since)}
    \item \texttt{address(user\_id, street, city, state)}
\end{itemize}

Here, the \texttt{friend} table implements a many-to-many relationship between users and their friends through foreign keys. In a document-oriented system, the same conceptual information may be represented differently---we might have a single \texttt{user} collection where each document contains an embedded array of \texttt{friends} and an embedded \texttt{address} subdocument.

In a graph database, \texttt{user} and \texttt{address} would be represented as node labels, while \texttt{FRIEND} and \texttt{USER\_ADDRESS} relationships would connect the nodes to each other.

Although these representations differ structurally, they capture the same conceptual semantics. Within the schema category, \texttt{User}, \texttt{Friend}, and \texttt{Address}, along with their properties, are modeled as objects, while links between them are represented as morphisms. Concrete design choices---such as join tables, document embeddings, or graph edges---correspond to alternative mapping configurations defined over the same canonical schema.

\begin{figure}[h]
    \centering
    \includegraphics[width=\textwidth]{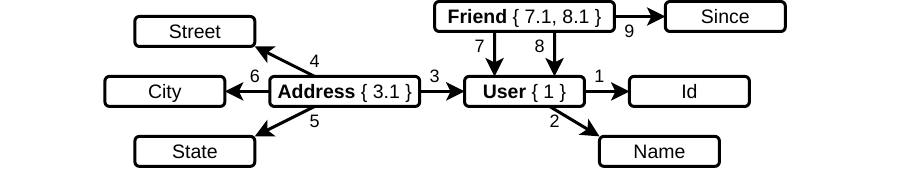}
    \caption{Example schema category for the user-friend-address model.}
    \label{fig:drawio/example-schema}
\end{figure}

The example schema category is depicted in Figure~\ref{fig:drawio/example-schema}. Note the numbers next to morphisms, which are called \emph{signatures} and whose purpose is to uniquely identify morphisms. Next, the numbers in curly braces represent identifiers of objects. For example, the \texttt{User} object is identified by the property accessible through the morphism with signature \texttt{1}, i.e., the \texttt{Id} property. \texttt{Address} is identified by a \emph{composite morphism} \texttt{3.1} which ultimately leads to the \texttt{Id} property. Lastly, \texttt{Friend} is identified by a composite key of two user identifiers.

\subsection{Schema and Instance Categories}

The \emph{schema category} captures structural definitions independent of a particular database system. Relational tables, document collections, or graph labels are treated uniformly as objects within the same structure. Structural differences---such as normalization, embedding, or edge projection---are expressed through morphism configurations.

\begin{figure}[h]
    \centering
    \includegraphics[width=\textwidth]{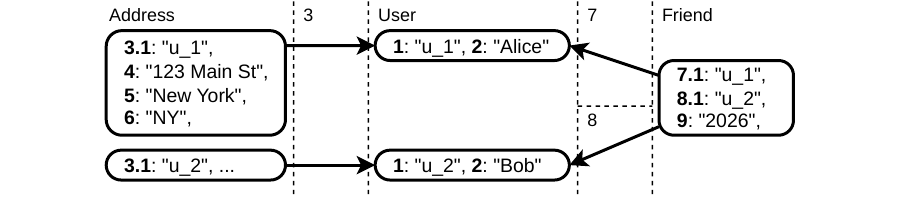}
    \caption{Example instance category for the schema category in Figure~\ref{fig:drawio/example-schema}. The dashed lines separate the domains of the instance objects/morphisms.}
    \label{fig:drawio/example-instance}
\end{figure}

The \emph{instance category} mirrors the schema category but contains concrete data. An example of such structure is shown in Figure~\ref{fig:drawio/example-instance}. There are five object records, each represented as a tuple (signature, value). These records are divided among domains of three instance objects (\texttt{User}, \texttt{Friend}, and \texttt{Address}).

There are also four morphism records, represented as tuples consisting of two object records. Each of the morphism records belongs to the domain of one of the three morphisms (\texttt{3}, \texttt{7}, and \texttt{8}).

\subsection{Mappings and Structural Transformations}

Mappings connect system-specific schemas with the unified schema category. They define how \emph{kinds} (relational tables, document collections, or graph nodes and relationships) correspond to canonical objects and morphisms.

For instance, a relational join table mapping would specify how foreign keys relate to morphisms in the schema, while a document embedding mapping would indicate which properties are nested within others.
In the running example, choosing a relational join table, document embedding, or graph edge representation corresponds to defining different mappings over the same conceptual schema. Each mapping yields a structurally distinct dataset while preserving logical semantics.

\begin{figure}[h]
    \centering
    \includegraphics[width=\textwidth]{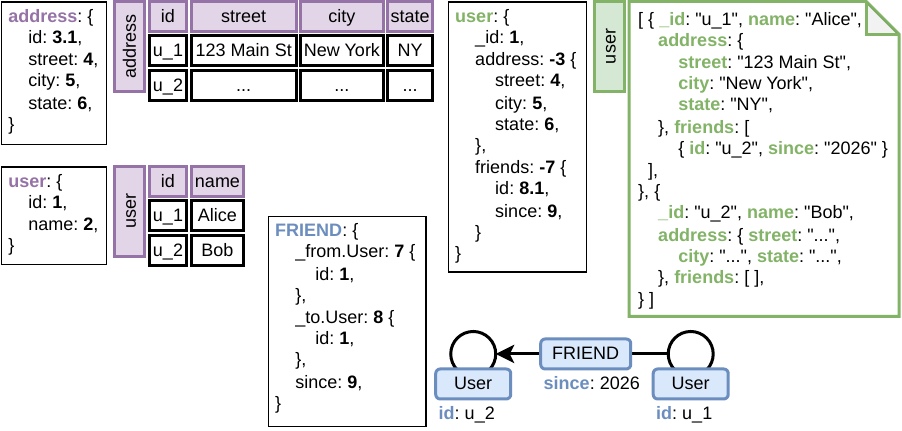}
    \caption{Example mapping from the schema category in Figure~\ref{fig:drawio/example-schema} to various kinds in relational (purple), graph (blue), and document (green) models.}
    \label{fig:drawio/example-mapping}
\end{figure}

Examples of such mappings are shown in Figure~\ref{fig:drawio/example-mapping}. Each mapping consists of a JSON-like structure where attribute names correspond to properties of the given kind and values correspond to signatures of morphisms in the schema category. For example, object \texttt{User} is mapped to a relational table \texttt{user} with attributes \texttt{id} and \texttt{name} corresponding to morphisms with signatures \texttt{1} and \texttt{2}, respectively. The same object is also mapped to a document collection \texttt{user} which, however, contains some additional attributes (nested document \texttt{address} and an array of nested documents \texttt{friends}). Relational model cannot contain nested attributes, so \texttt{Address} has to be mapped to a separate table.

This explicit mapping mechanism enables controlled structural variation. For benchmarking, this is critical: multiple structural realizations of the same dataset can be generated and evaluated under identical logical semantics. Structural evolution can be interpreted as a sequence of mapping modifications applied to the schema category and propagated to the instance category.

\new{
\subsection{Formalization}

\def\os{O_{\mathcal{S}}}
\def\ms{M_{\mathcal{S}}}

Formally, a \emph{schema category} $\mathcal{S}$ is a category $\mathcal{S} = (\os, \ms, \circ)$, where $\os$ is a set of objects, $\ms$ is a set of morphisms between the objects, and $\circ$ is the composition operation.

Similarly, we can define an instance category $\mathcal{I}$ as a category of \emph{instance objects} and \emph{instance morphisms} that mirror the structure of the schema category and contain concrete data. An \emph{instance} is then a functor $\mathrm{inst}: \mathcal{S} \to \mathcal{I}$, mapping schema category to instance category while preserving the structure defined by the morphisms. 

Finally, a \emph{mapping} from the schema category $\mathcal{S}$ to a system-specific logical representation $\mathcal{C}$ is defined as a functor $M: \mathcal{S} \to \mathcal{C}$, where $\mathcal{C}$ is the category representing the system-specific schema. The functor preserves the structure of the schema category while translating it into the target representation.

We use category theory as a formal foundation for our framework, but the practical implementation abstracts away from these details. This allows us to focus on the engineering aspects of benchmark construction while ensuring a rigorous theoretical basis for structural representation and transformation.
}

\new{
\subsection{Implications for Benchmark Construction}

In this work, we do not claim full semantic equivalence in the sense of arbitrary query equivalence across different data models. Instead, we use the term \emph{semantic alignment} to denote that generated variants are derived from the same canonical schema category and instance category, preserve the same selected entity identities, properties, and references, and are produced by explicit mappings recorded as part of the benchmark configuration. Thus, two variants may differ structurally, for example by normalization, embedding, or graph projection, but their corresponding elements remain traceable to the same canonical objects and morphisms. This notion is sufficient for the benchmark engineering objective of this paper: evaluating how alternative representations of the same conceptual data affect selected workloads.
}

The combination of unified schema representation, instance-level alignment, and explicit mappings enables configurable benchmark engineering. Instead of evaluating systems using a single fixed schema, multiple structural variants and evolution scenarios can be systematically generated and reproduced.

This capability is particularly important in heterogeneous information systems, where data models coexist, and structural redesign is common. By grounding benchmark construction in a unified abstraction, the framework allows rigorous evaluation of representation choices and transformation strategies across evolving multi-model environments.

The following section describes how these conceptual foundations are operationalized within the proposed benchmarking framework.

%% file: chapters/chap04.tex
\section{Design of the Evolution-Aware Benchmarking Framework}
\label{sec:tool}

Based on the conceptual foundations introduced in the previous section, we now present the design of an evolution-aware benchmarking framework for heterogeneous information systems and its implementation in the \emph{TransforMMer} platform. The framework operationalizes unified multi-model schema representation and transformation-driven dataset generation, enabling systematic construction of benchmark datasets across heterogeneous data models.

While \emph{TransforMMer} has been previously introduced in a vision paper~\cite{holubova2025reshaping} and a demonstration paper~\cite{bartik2025transformmer}, those works focused primarily on conceptual motivation and prototype demonstration. In contrast, this paper positions the framework as an infrastructure for reproducible and evolution-aware benchmarking and provides a detailed description of its design principles, workflow, and architectural organization.

\subsection{Design Rationale}

The framework is designed to address three core requirements of benchmarking in heterogeneous information systems:

\begin{itemize}
    \item \textbf{Unified structural representation:} Heterogeneous datasets originating from relational, document, or graph models must be represented in a model-independent manner.
    \item \textbf{Controlled evolution:} Benchmark datasets should support structural redesign and schema evolution scenarios that reflect real-world system changes. \new{This encompasses both (i)~conceptual schema changes propagated consistently to all logical representations and queries, and (ii)~workload-driven adaptations of logical representations that preserve the underlying conceptual schema.}
    \item \textbf{Reproducibility:} Alternative structural variants must be generated in a controlled and repeatable way to enable systematic comparison.
\end{itemize}

These requirements motivated the separation of the benchmarking process into three stages: schema inference, unified schema construction, and transformation-driven dataset generation. This separation allows structural modifications to be introduced at a conceptual level and propagated consistently across multiple target representations.

\subsection{Benchmark Generation Workflow}
\label{sec:workflow}

The framework follows a structured pipeline that transforms heterogeneous input data into reproducible multi-model benchmark datasets. The workflow consists of three main stages:

\begin{enumerate}
    \item \textbf{Schema inference} from raw input data,
    \item \textbf{Unified schema construction and refinement},
    \item \textbf{Data transformation and generation} for selected target systems.
\end{enumerate}

\begin{figure}[h]
    \centering
    \includegraphics[width=\textwidth]{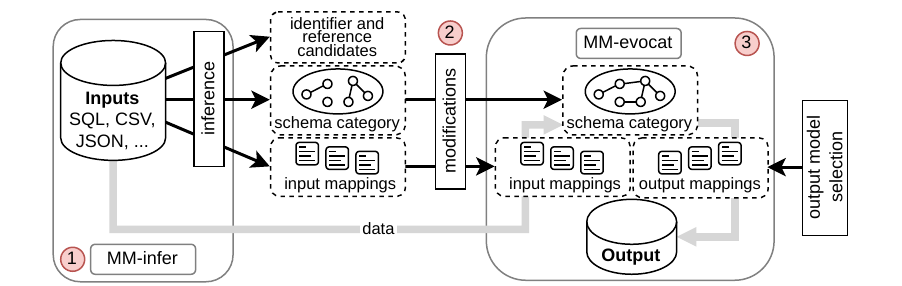}
    \caption{Benchmark generation workflow from heterogeneous input data to multi-model benchmark outputs. Adapted from~\cite{bartik2025transformmer}. \new{See Figure~\ref{fig:drawio/example-workflow} for a concrete example of the workflow stages.}}
    \label{fig:drawio/example-architecture}
\end{figure}

Figure~\ref{fig:drawio/example-architecture} illustrates the overall workflow.

\begin{figure}[h]
    \centering
    \includegraphics[width=\textwidth]{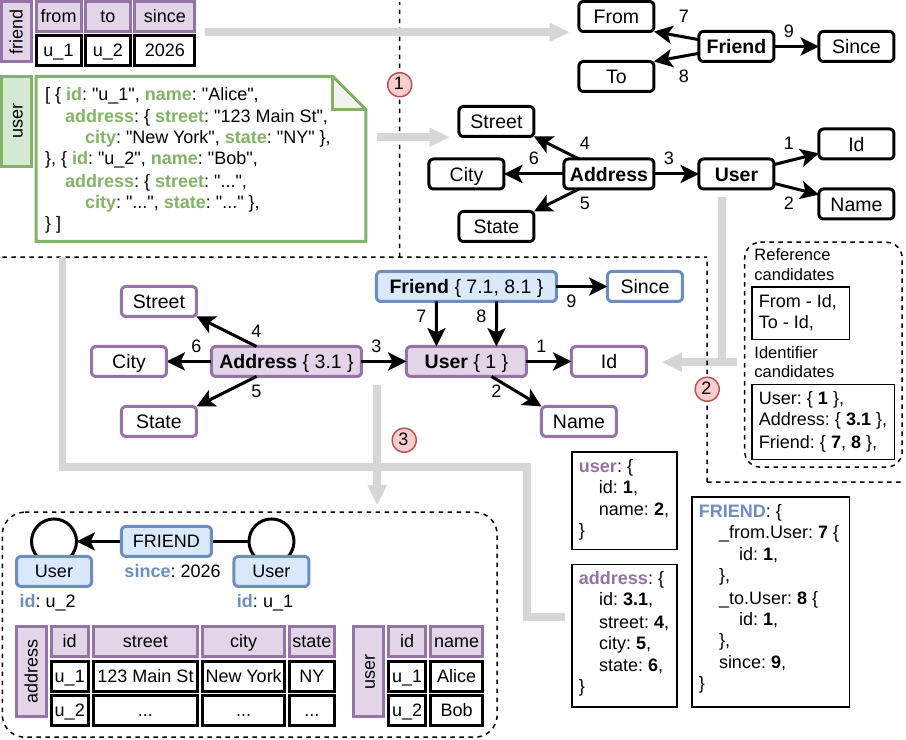}
    \caption{Inputs (JSON and CSV data), intermediate artifacts (schema category components, reference and identifier candidates), and outputs (unified schema, relational and graph data) of the user-friend-address example. The step numbers correspond to the workflow stages from Figure~\ref{fig:drawio/example-architecture}.}
    \label{fig:drawio/example-workflow}
\end{figure}

To illustrate the workflow, we continue with the dataset describing \textit{Users} and \textit{Friends} (see Figure~\ref{fig:drawio/example-workflow}). Assume the input is provided in a semi-structured format (e.g., JSON), where user records contain nested addresses. Another input (e.g., CSV) then contains the friendships. The framework first performs schema inference, extracting structural elements such as entities (i.e., \texttt{User}, \texttt{Address}, and \texttt{Friend}), attributes (e.g., \texttt{Id}), reference candidates (e.g., from \texttt{From} to \texttt{Id}), identifier candidates (e.g., \texttt{User} $\rightarrow$ \texttt{Id}), and mappings between the input data models and the inferred schema category components.

By accepting the candidates, the user merges the schema category components into one and defines object identifiers. Candidates can also be rejected, or created manually. The refined schema is displayed in Figure~\ref{fig:drawio/example-schema}. At this level, the conceptual relationship between users, friends, and addresses is represented independently of whether it will later be realized as a join table, an embedded collection, or a graph edge. In the background, the framework updates the inferred mappings to reflect the changes to the schema category. We can use these mappings to create an instance category that represents the data in a model-independent way (see Figure~\ref{fig:drawio/example-instance}).

Finally, alternative structural mappings are generated for relational, document, or graph database systems. \old{Each mapping yields a structurally distinct yet semantically equivalent dataset variant derived from the same conceptual schema.}\new{Each mapping yields a structurally distinct yet semantically aligned dataset variant derived from the same conceptual schema and traceable to the same canonical entities, properties, and references.} In this example, the user decides to map the \texttt{User} and \texttt{Address} entities to relational tables, while representing the \texttt{Friend} entity as graph edges. These mappings are then used to export data from the instance category into the target systems.

The workflow produces reusable intermediate artifacts, including inferred schema descriptions, unified schema representations, and explicit transformation mappings. These artifacts enable reproducible modification and regeneration of benchmark configurations under alternative representation strategies.

\paragraph{Schema Inference}

The first stage derives structural information from input datasets. Input data may originate from heterogeneous sources such as CSV exports, JSON documents, or relational dumps. The framework extracts entities, attributes, nested structures, and references and gathers statistical characteristics, including attribute frequencies and candidate keys.

Inference functionality is implemented using the MM-infer component~\cite{Koupil2022a}, which performs distributed processing based on Apache Spark~\cite{Zaharia2010Spark}. Local structural descriptions are merged into a global representation that serves as the basis for unified schema construction.

\paragraph{Unified Schema Construction}

The inferred structural representation is transformed into a unified schema category that captures entities and relationships independently of any specific data model. This representation serves as a stable structural reference for defining benchmark configurations and exploring alternative mappings.

An important motivation for constructing the unified schema is that genuinely multi-model datasets are rarely available in practice. Most real-world data is accessible in a single representation (e.g., CSV, JSON, or relational exports), and heterogeneous multi-model versions must be constructed manually. By transforming inferred single-model structures into a unified schema, the framework establishes a conceptual representation from which multiple model-specific variants can be systematically generated. In this way, the unified schema acts as a bridge between single-model input data and reproducible multi-model benchmark datasets.

At this stage, structural refinements may be introduced. These refinements correspond to realistic redesign operations observed in evolving information systems, including:
\begin{itemize}
    \item normalization or denormalization,
    \item attribute embedding,
    \item schema enrichment through integration of additional data sources,
    \item partitioning of entities across models.
\end{itemize}

Such refinements can be applied either immediately after schema inference or in later iterations of the workflow. Once the unified schema has been established, subsequent benchmark variants can be generated by modifying mappings and structural configurations rather than re-inferring structure from raw data. This enables simulation of schema and system evolution over time while preserving comparability across benchmark versions.

By introducing refinements at the conceptual level, the framework ensures that alternative target representations remain comparable while reflecting controlled structural variation.

\paragraph{Data Transformation and Generation}

In the final stage, refined schemas and mappings are used to generate datasets for selected database systems. Depending on the configuration, the framework produces relational tables, document collections, graph structures, or hybrid multi-model outputs.

Transformations are executed using the MM-evocat framework~\cite{Koupil2022c}, which manages mapping execution and schema evolution operations. The generated datasets represent alternative structural realizations of the same conceptual data and serve as benchmark inputs for comparative evaluation.

\subsection{Framework Architecture}

The framework's architecture mirrors the three-stage workflow and comprises corresponding layers: inference, unified schema construction, and transformation (see Figure~\ref{fig:drawio/example-architecture}).
The inference layer extracts structural information from heterogeneous input sources. The schema construction layer builds and refines the unified schema representation. The transformation layer generates target-specific datasets and manages versioned benchmark outputs. Intermediate artifacts are preserved to ensure reproducibility and facilitate controlled reconfiguration.
This layered architecture supports modular extensibility while maintaining a coherent transformation pipeline.

\subsection{Benchmark Engineering Capabilities}

The framework enables systematic construction of benchmark datasets that reflect alternative structural representations and evolution scenarios. Since genuinely heterogeneous multi-model datasets are rarely available in practice, the workflow starts from single-model inputs and derives a unified conceptual schema from which multiple representation variants can be generated. Instead of relying on a single static dataset, the framework produces a family of benchmark variants derived from the same conceptual data through controlled transformations and mapping configurations.

\paragraph{Dynamic Benchmarks}

Dynamic benchmarks reflect structural or content changes over time and allow evaluation under evolving conditions. After the initial schema inference and unified schema construction, subsequent benchmark versions can be generated by modifying mappings and applying structural refinements, rather than reprocessing raw data. The framework supports dynamic benchmarking through versioned dataset generation and controlled, conceptual-level schema transformations.
\new{Crucially, structural modifications may also be introduced exclusively at the logical representation level---by modifying mappings while keeping the schema category fixed---thereby enabling systematic exploration of workload-driven adaptation scenarios in which changing access patterns, rather than schema modifications, motivate structural redesign.}

\paragraph{Real-World Dataset Evolution}

When real-world datasets evolve, updated versions can be processed through the same pipeline while preserving transformation configurations and mappings. Because the unified schema serves as a stable structural reference, new dataset versions can be aligned with existing benchmark variants. This ensures comparability across dataset versions and supports longitudinal evaluation of architectural decisions under changing data conditions.

\paragraph{Simulated Schema Evolution}

If only static data is available, evolution scenarios can be simulated directly at the unified schema level. Structural refinements or alternative mappings can be introduced without repeating schema inference, enabling controlled creation of benchmark variants that represent alternative structural configurations. In this way, the framework supports experimentation with redesign operations such as embedding, partitioning, or enrichment, even in the absence of naturally evolving data.

To support reproducibility, generated datasets are stored together with mappings, transformation specifications, and version metadata describing the applied refinements. This enables systematic comparison of benchmark configurations and repeatable experimental evaluation across iterations.

By treating benchmark construction as a configurable, versioned engineering process rather than a fixed dataset design task, the framework supports the systematic evaluation of evolving heterogeneous information systems and enables reproducible exploration of representation-level and architectural design decisions.

%% file: chapters/chap05.tex
\section{Evaluation through Evolution Scenarios in Information Systems}
\label{sec:experiments}

This section evaluates the proposed evolution-aware benchmark engineering approach implemented in \emph{TransforMMer}. Rather than evaluating database engines in isolation, we interpret each experiment as an information systems design scenario. The objective is to analyze how alternative architectural and representation decisions shape system behavior under comparable workloads and data conditions. Consequently, the experiments are not intended to establish a general ranking of database technologies. Instead, we use controlled benchmark scenarios to demonstrate that (i) comparable benchmark variants can be generated across heterogeneous data models, and (ii) representation choices and evolution-like redesign steps (e.g., embedding, enrichment, partitioning) lead to measurable and explainable differences in observed query costs. From an Information Systems perspective, the scenarios should be interpreted as controlled design experiments rather than database benchmarks. They illustrate how architectural and representation decisions influence the behavior of evolving heterogeneous systems.

The experiments are organized as \emph{benchmark construction scenarios}. Each scenario starts from the same underlying data and applies a specific transformation or representation decision. This mirrors the practice of realistic information systems, where data representations are incrementally adapted in response to evolving requirements and workload pressure.

\subsection{Inputs}

The datasets were selected to be representative of real-world data and to cover a range of structural characteristics relevant to heterogeneous benchmarking.

First, publicly available datasets are most commonly distributed as CSV or JSON. We therefore assume that typical \emph{TransforMMer} users start from these formats, and we adapted our input selection accordingly.

Second, we prefer datasets that contain non-trivial structure, including nested objects and references. This supports meaningful transformations and enables workloads reflecting realistic access patterns (joins, multi-hop traversals, aggregation).

Finally, we also considered well-known or easy-to-understand datasets, which support clarity and reproducibility.

\paragraph{Yelp Dataset}

The Yelp dataset\footnote{\url{https://business.yelp.com/data/resources/open-dataset/}} is widely used as it provides well-structured real-world data for analytical tasks. It includes businesses, reviews, users, check-ins, and attributes, thus capturing consumer interactions and feedback. The data is provided in JSON format, with each file containing a different type (e.g., \textit{business.json}, \textit{review.json}, \textit{user.json}). The subset used in our experiments is approximately 28 MB.

\begin{figure}
    \centering
    \includegraphics[width=\textwidth]{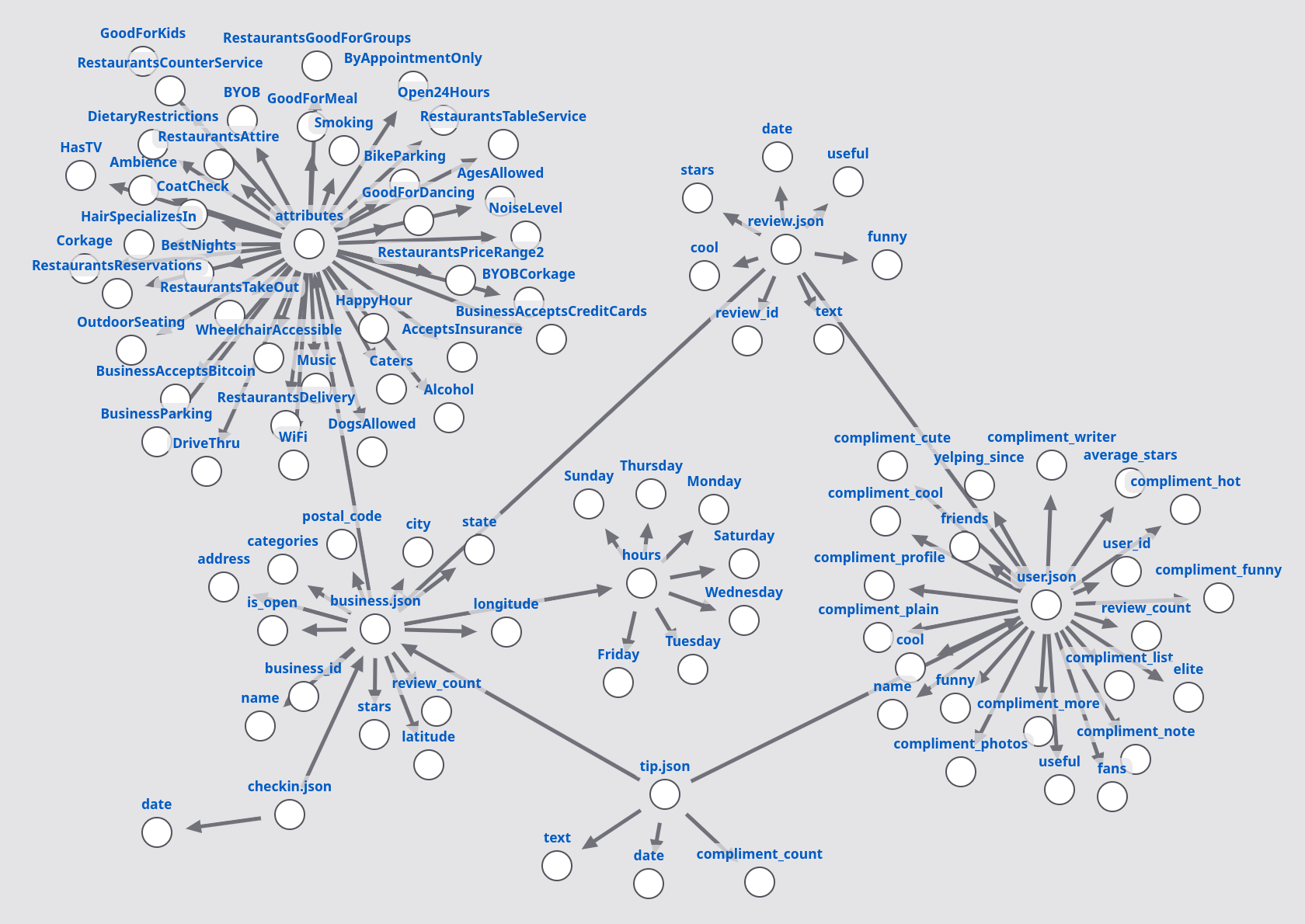}
    \caption{Schema category of the Yelp dataset inferred by \emph{TransforMMer}.}
    \label{fig:sc/yelp}
\end{figure}

Figure~\ref{fig:sc/yelp} shows the inferred schema category. It contains five kinds corresponding to the five input files: \textit{business.json}, \textit{user.json}, \textit{review.json}, \textit{tip.json}, and \textit{checkin.json}. The inferred schema category exhibits a complex structure with many attributes and links between kinds, which makes it suitable for studying alternative representations and their impact on query execution.

\paragraph{BibleData Dataset}

\begin{figure}
    \centering
    \includegraphics[width=\textwidth]{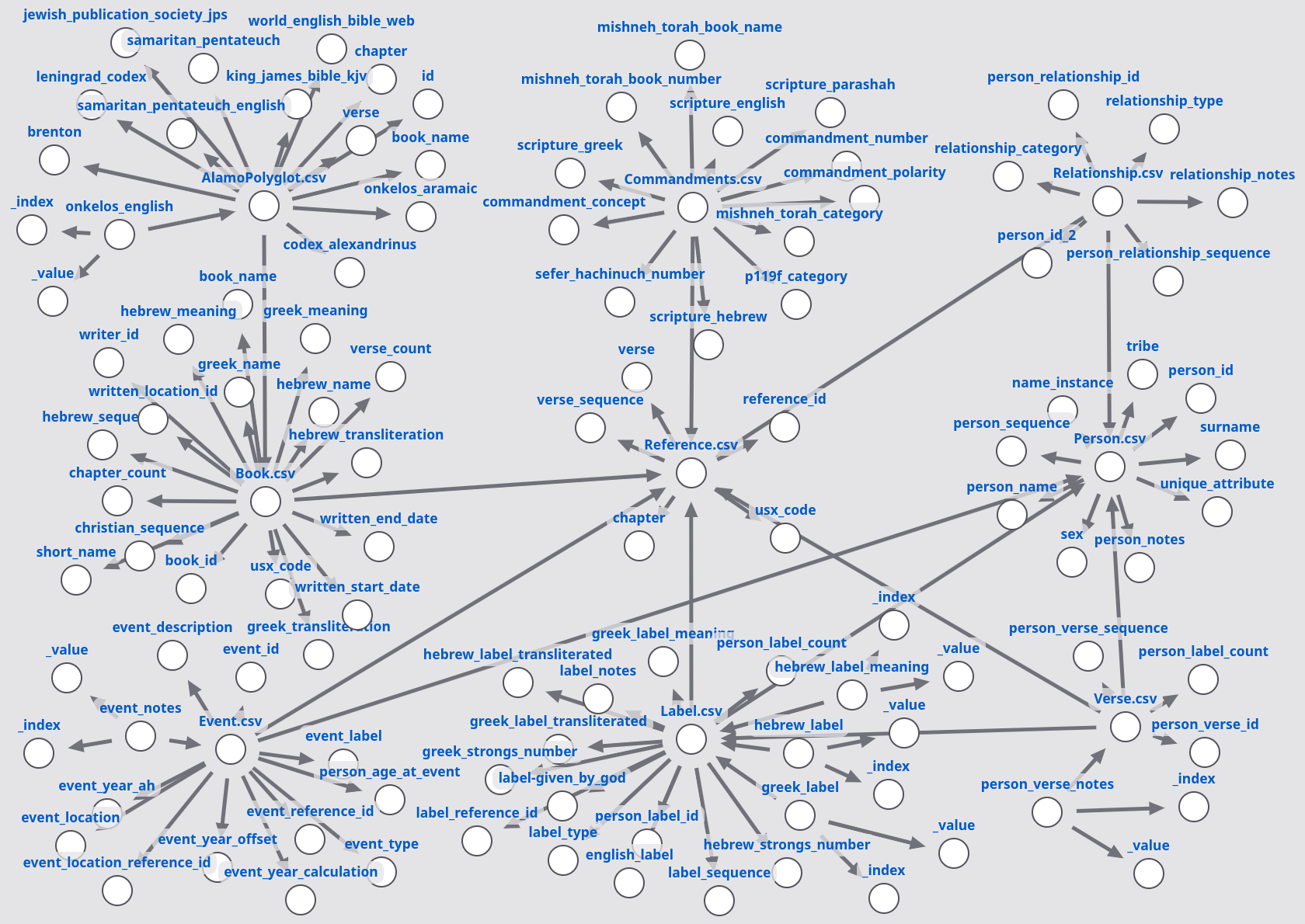}
    \caption{Schema category of the BibleData dataset  inferred by \emph{TransforMMer}.}
    \label{fig:sc/bibledata}
\end{figure}

The BibleData dataset\footnote{\url{https://www.kaggle.com/datasets/bradystephenson/bibledata}} is a large and complex collection of structured information about Bible texts, translations, and metadata in CSV format. It includes persons, relationships, verses, labels, and events, with rich interconnections. This structure is reflected in the inferred schema category in Figure~\ref{fig:sc/bibledata}.

\new{This dataset is not used as an example of a rapidly evolving application domain. It is used as a structurally rich, relationship-centric dataset that allows us to test graph-oriented and hybrid representations under controlled conditions.}

The full dataset contains 18 files; to keep the experiments manageable, we used 9 files (9 types), totaling approximately 26~MB.

\paragraph{US Cities Data}

\begin{figure}
    \centering
    \includegraphics[width=0.4\textwidth]{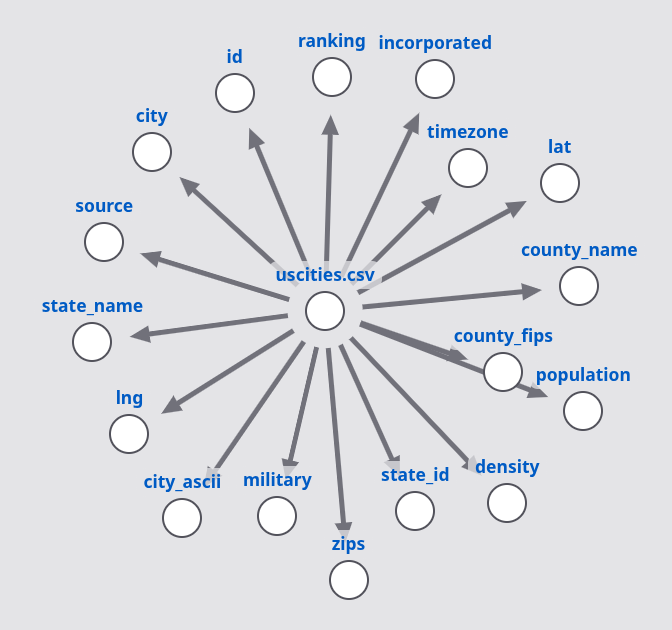}
    \caption{Schema category of the US Cities data inferred by \emph{TransforMMer}.}
    \label{fig:sc/uscities}
\end{figure}

Finally, we used a simple US cities dataset\footnote{\url{https://simplemaps.com/data/us-cities}} (nearly 3{,}000~KB) provided as CSV. This dataset is used later to enrich the Yelp dataset. Figure~\ref{fig:sc/uscities} shows its schema category.

\subsection{Experimental Setup}

The experiments explore scenarios that combine (i) structural transformations, (ii) alternative representation choices, and (iii) query workloads executed over the generated datasets. The purpose is to demonstrate that \emph{TransforMMer} can generate comparable benchmark datasets across models and that representation choices---including evolution-like redesign steps such as embedding or partitioning---have a measurable impact on observed query costs.

\paragraph{Queries}

All benchmark queries were manually created and translated into the native query languages of the target systems. While the current version of \emph{TransforMMer} focuses on data transformation, we plan to extend the approach to support the transformation of datasets and their associated queries. In this direction, our team is actively working on a generic query language and internal representation logic~\cite{koupil_universal_2025}.

\paragraph{Measurements}

Query performance measurements were conducted as follows:
\begin{itemize}
    \item Database systems were executed in \textbf{Docker} containers.
    \item Only the relevant database container was running during measurements.
    \item Before each set of measurements, the container was \textbf{restarted} to reduce caching effects.
    \item Query execution was automated using \textbf{Python scripts}.
    \item Each query was executed \textbf{60 times}; the \textbf{last 50 runs} were used to compute the \textbf{average execution time}. The first 10 runs were excluded as a warm-up phase.
    \item We ensured that each database system contained exactly \textbf{the same dataset}. This was automatically enforced by \emph{TransforMMer}.\footnote{Different systems may store and access data differently; such differences are inherent and may influence performance.}
\end{itemize}

\new{
\paragraph{Semantic Alignment Checks}
To substantiate comparability of the generated variants, we checked semantic alignment at the level of the benchmark configuration. The checks included preservation of object identifiers, cardinalities of mapped object sets, reference consistency, and agreement of selected validation queries formulated over the canonical schema and translated to the target representation. These checks do not establish general semantic equivalence between database models; rather, they document that the generated variants are comparable for the explicitly defined benchmark workload.}

\paragraph{Technical Details}

Hardware specifications:
\begin{itemize}
    \item \textbf{CPU:} Intel(R) Core(TM) i7-8565U CPU @ 1.80GHz, 4 cores / 8 threads
    \item \textbf{RAM:} 16 GB DDR4 @ 2400 MHz
    \item \textbf{Operating System:} Windows 11, Version 24H2 (OS Build 26100.3775)
    \item \textbf{Execution Environment:} Native execution on a local machine; database systems were run in Docker containers.
\end{itemize}
Software environment:
\begin{itemize}
    \item \textbf{Docker Version:} Docker Version 24.0.6
    \item \textbf{Python Version:} Python 3.12.2
\end{itemize}
Database systems:
\begin{itemize}
    \item MongoDB
    \begin{itemize}
        \item \textbf{Name and Version:} MongoDB 8.0.8
        \item \textbf{Docker Image:} \texttt{mongo:8.0.8}
    \end{itemize}
    \item PostgreSQL
    \begin{itemize}
        \item \textbf{Name and Version:} PostgreSQL 17.4
        \item \textbf{Docker Image:} \texttt{postgres:17.4}
    \end{itemize}
    \item Neo4j
    \begin{itemize}
        \item \textbf{Name and Version:} Neo4j 2025.03.0
        \item \textbf{Docker Image:} \texttt{neo4j:5.15}
    \end{itemize}
\end{itemize}

\paragraph{Setup Limitations and Interpretation}

The setup was controlled to keep the measurements comparable between systems. Nevertheless, certain limitations can influence the stability. The experiments were run on a personal Windows machine using Docker containers. In Windows, Docker typically relies on a virtualized Linux environment, which can introduce overhead and variability in resource scheduling compared to native Linux or dedicated servers~\cite{felter2015performance}. Desktop background processes, thermal throttling, or variable CPU clock speeds can also contribute to noise.

\begin{figure}
    \centering
    \includegraphics[width=0.9\textwidth]{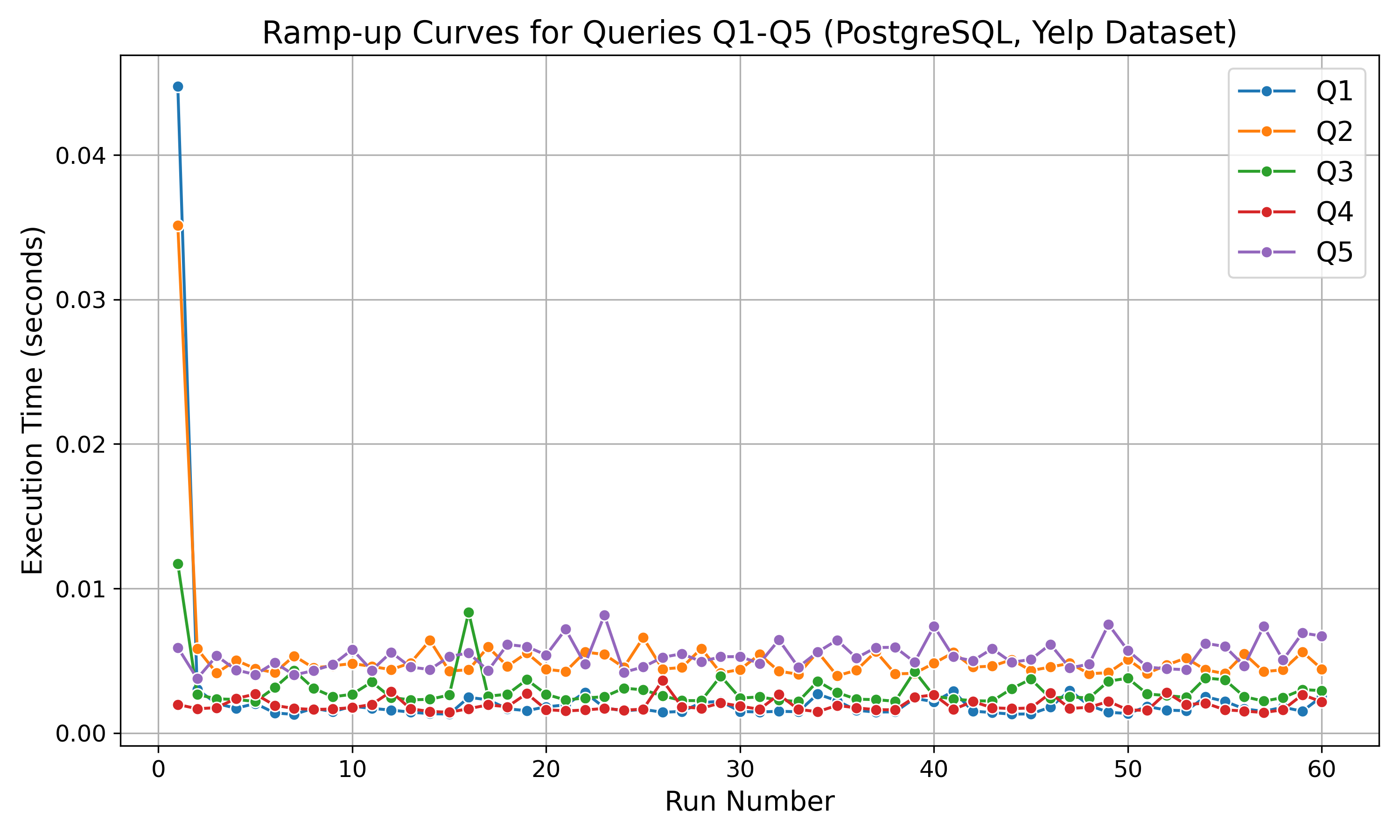}
    \caption{Ramp-up curves for the first five queries (Q1--Q5) executed on PostgreSQL with the Yelp dataset.}
    \label{fig:ramp-up}
\end{figure}

To reduce these effects, containers were restarted before measurements, no other workloads were active, and we used a warm-up phase. Figure~\ref{fig:ramp-up} illustrates the stabilization behavior over 60 repetitions for the first five queries of the Yelp experiment on PostgreSQL (Section~\ref{sec:query-yelp}). The first runs show elevated execution times; later runs stabilize. We excluded the first 10 runs for robustness between experiments. Therefore, the reported results are internally consistent and suitable for comparing \emph{relative trends} caused by structural transformations.

However, absolute execution times should not be interpreted as definitive deployment benchmarks. The primary objective is to observe how alternative representations and evolution-like redesign steps influence performance within a reproducible experimental setting.

\subsection{Benchmark Construction Scenarios}
\label{sec:benchmarks}

This subsection presents benchmark scenarios constructed using \emph{TransforMMer}. Each scenario starts from the same underlying data but applies a controlled transformation or representation decision. In this way, the experiments emulate typical evolution steps in information systems, such as denormalization/normalization, embedding, enrichment, or hybrid partitioning.

\subsubsection{Yelp: Query Performance under Alternative Representations}
\label{sec:query-yelp}

We start with the Yelp dataset and evaluate how different representations of the same data influence query costs in relational and document settings.

Table~\ref{tab:yelp-queries} lists the query set. Although most queries were created for this work, \textbf{Q2} and \textbf{Q3} are inspired by existing studies to maintain comparability and realism. Query \textbf{Q2} was inspired by Zhang~\cite{zhang2014yelp}, where Yelp data is loaded into PostgreSQL~\cite{postgresql} and explored using joins and aggregations. Query \textbf{Q3} is inspired by both Zhang~\cite{zhang2014yelp} and Alam et al.~\cite{alam2021yelp}. The remaining queries cover patterns from filtering and sorting to join-intensive and multi-hop style access.

\begin{table}[h]
\footnotesize
    \caption{Experimental Queries for Yelp dataset}
    \label{tab:yelp-queries}
    \centering
    \begin{tabular}{@{}cp{4.5cm}p{2cm}p{2.5cm}p{2cm}@{}}
      \toprule
      \textbf{Query} & \textbf{Query Description} & \textbf{Type} & \textbf{Inspired by} & \textbf{Note} \\
      \midrule
      \textbf{Q1} & Find top 10 businesses in New Orleans with rating $\geq 4.5$ and more than 100 reviews. & Filter + Sort & New & Classic benchmark \\
      \textbf{Q2} & Find users who have reviewed at least 2 businesses in the same city. & Group + Join & Zhang 2014~\cite{zhang2014yelp} & User preference patterns \\
      \textbf{Q3} & Count how many businesses saw a drop in average rating from 2019 to 2021. & Temporal Aggregation & Zhang 2014~\cite{zhang2014yelp}, Alam et al. 2021~\cite{alam2021yelp} & Rating trend signal \\
      \textbf{Q4} & For a given user, find all other users who reviewed the same businesses. & Join & New & Multi-hop style query \\
      \textbf{Q5} & Find the average star rating per state per year. & Aggregation & New & Time-series aggregation \\
      \textbf{Q6} & For each city, find the user with the highest average review stars. & Group + Max & New & Grouping and ranking \\
      \textbf{Q7} & Find businesses frequently reviewed by the same group of users ($\geq$3). & Pattern & New & Join-heavy workload \\
      \bottomrule
    \end{tabular}
\end{table}

\begin{figure}
    \centering
    \includegraphics[width=\textwidth]{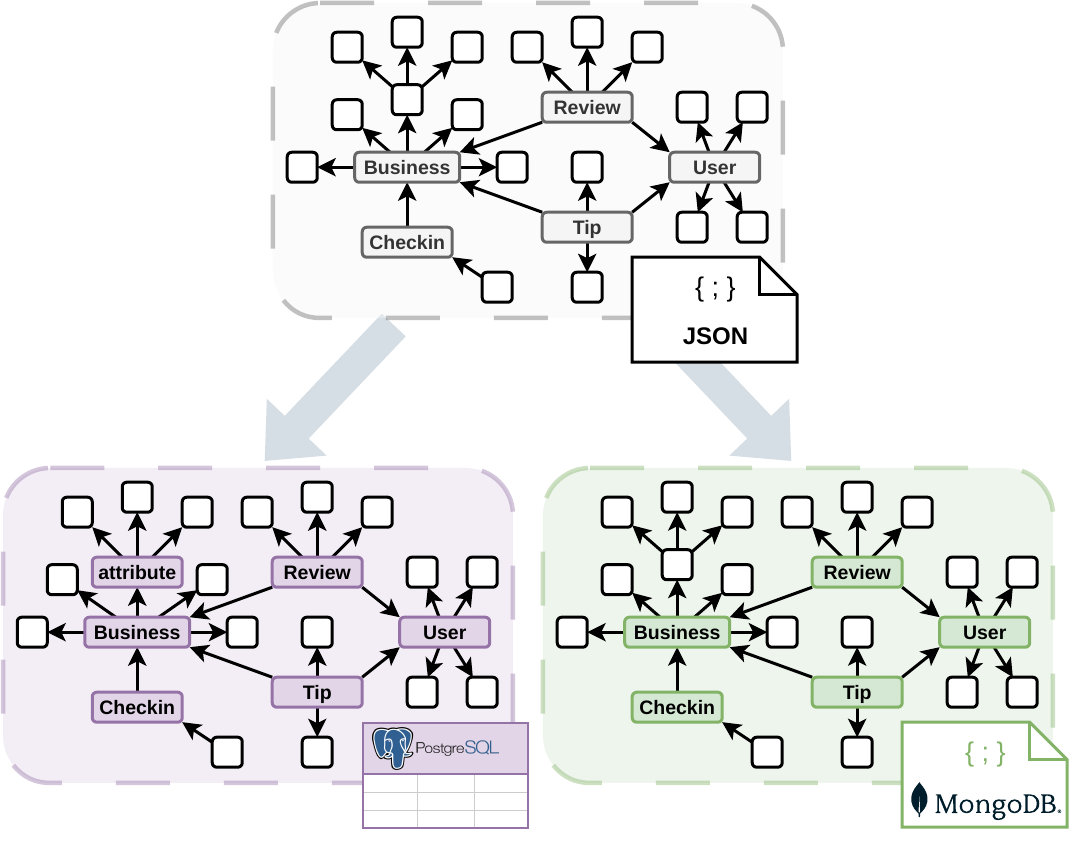}
    \caption{Transformations of the Yelp dataset to MongoDB and PostgreSQL}
    \label{fig:drawio/yelp-base}
\end{figure}

\paragraph{\textbf{Scenario A:} Alternative Persistence Architectures for the Same Information System}
\label{sec:mongo-vs-postgres}
This scenario represents an architectural decision between alternative persistence strategies for the same information system. We first construct two baseline representations: a document-oriented representation in MongoDB~\cite{mongodb} and a relational representation in PostgreSQL. Figure~\ref{fig:drawio/yelp-base} summarizes the transformation. In both cases, we transform the full schema category. For PostgreSQL, kinds with complex properties require decomposition into relationally manageable tables; this is illustrated for \textit{Business}, where nested \textit{attributes} are separated.

\begin{figure}
    \centering
    \includegraphics[width=0.9\textwidth]{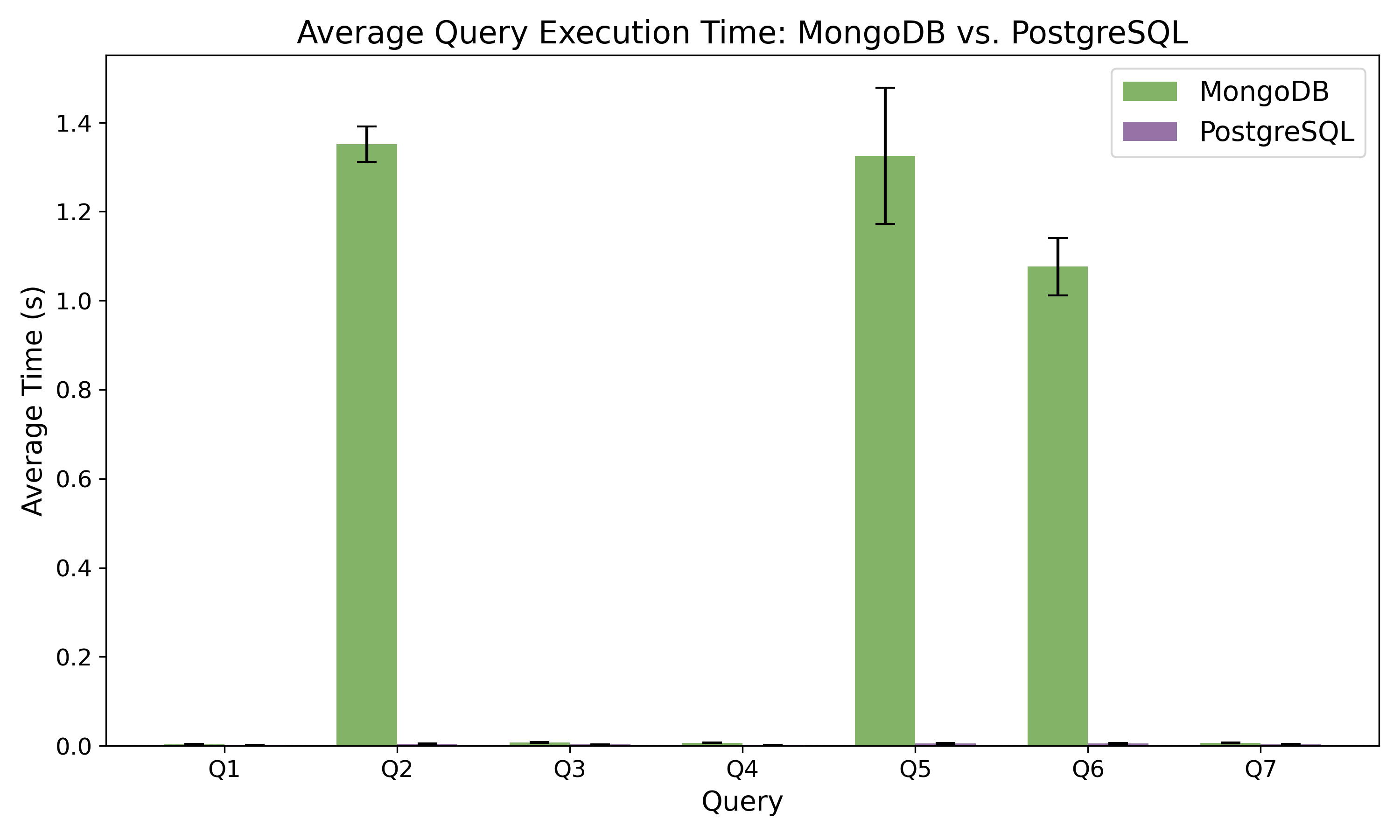}
    \caption{Average execution times of the queries from Table~\ref{tab:yelp-queries} on the Yelp dataset, comparing MongoDB and PostgreSQL. Error bars represent one standard deviation over the last 50 runs per query.}
    \label{fig:plot/yelp-base}
\end{figure}

\begin{figure}
    \centering
    \includegraphics[width=0.9\textwidth]{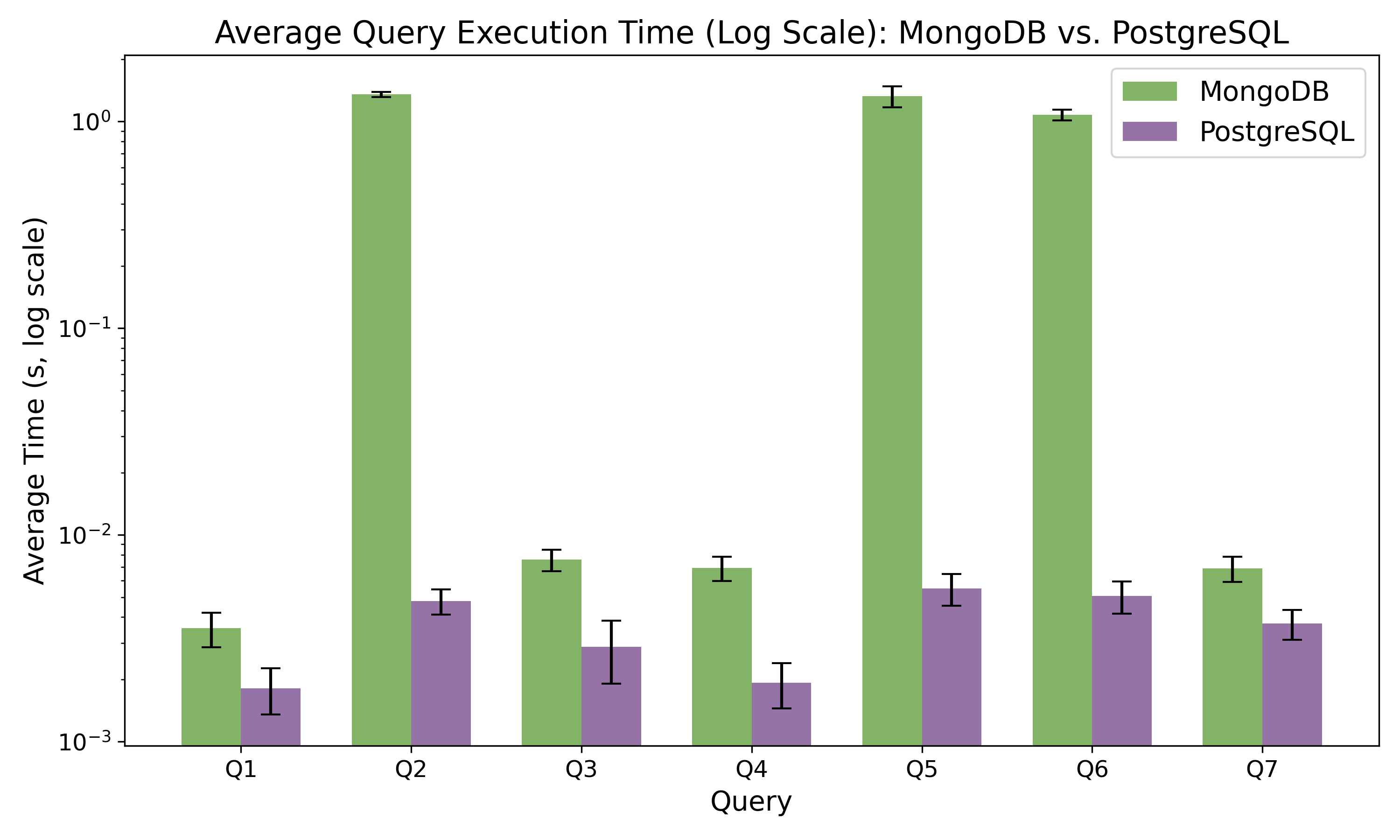}
    \caption{Average execution times of the queries from Table~\ref{tab:yelp-queries} on the Yelp dataset, comparing MongoDB and PostgreSQL. Error bars represent one standard deviation over the last 50 runs per query. Time is shown on a logarithmic scale.}
    \label{fig:plot/yelp-base-log}
\end{figure}

We execute the query set (Table~\ref{tab:yelp-queries}) in the native query languages and compare average execution times. Figures~\ref{fig:plot/yelp-base} and~\ref{fig:plot/yelp-base-log} show the results (linear and log scale). The baseline scenario shows that the same analytical intent can yield different computational costs depending on the representation choices and query mechanisms. In particular, join-like patterns implemented via \texttt{\$lookup} stages tend to be more expensive in this configuration, while PostgreSQL executes join-heavy aggregations using more mature join and aggregation strategies.

The two inspired queries (\textbf{Q2}, \textbf{Q3}) support comparability with previous work. Zhang~\cite{zhang2014yelp} uses similar join-heavy exploration in PostgreSQL~\cite{postgresql}. Alam et al.~\cite{alam2021yelp} emphasize year-over-year trends. Our scenario extends these analyzes by quantifying the cost of comparable analytical patterns across alternative representations.

We also note that query \textbf{Q4} required a preparatory step in MongoDB (precomputing the list of reviewed businesses), which was not included in the measured execution time and may underestimate the MongoDB cost. In PostgreSQL, the logic is executed within a single query.

Finally, while PostgreSQL outperforms MongoDB in these join-heavy workloads, MongoDB can outperform relational systems in other scenarios, particularly at scale and under horizontal execution strategies, as discussed by Makris et al.~\cite{makris2020mongodb}.

\paragraph{\textbf{Scenario B:} Representation-Level Redesign Driven by Workload Requirements}
\label{sec:mongo-embedded-vs-postgres}

\old{This scenario reflects the evolution of the system through the integration of additional data sources.}\new{This scenario reflects workload-driven structural evolution within a document-oriented information system.} We emulate a typical evolution step in document-oriented information systems: embedding selected data to reduce join-like operations. Based on the baseline scenario, we embed selected business and user attributes into \textit{Review}. This introduces redundancy, but reduces the number of repeated \texttt{\$lookup} stages. Such a redesign is common when specific query patterns become performance-critical.
\new{From a workload perspective, this scenario illustrates that a shift in access patterns---for instance, an increased emphasis on review-centric queries that aggregate or filter business and user attributes---can motivate a change in the logical representation while leaving the conceptual schema unchanged. The framework generates this adapted representation through mapping modifications alone, without requiring re-inference or schema reconstruction.}

\begin{figure}
    \centering
    \includegraphics[width=\textwidth]{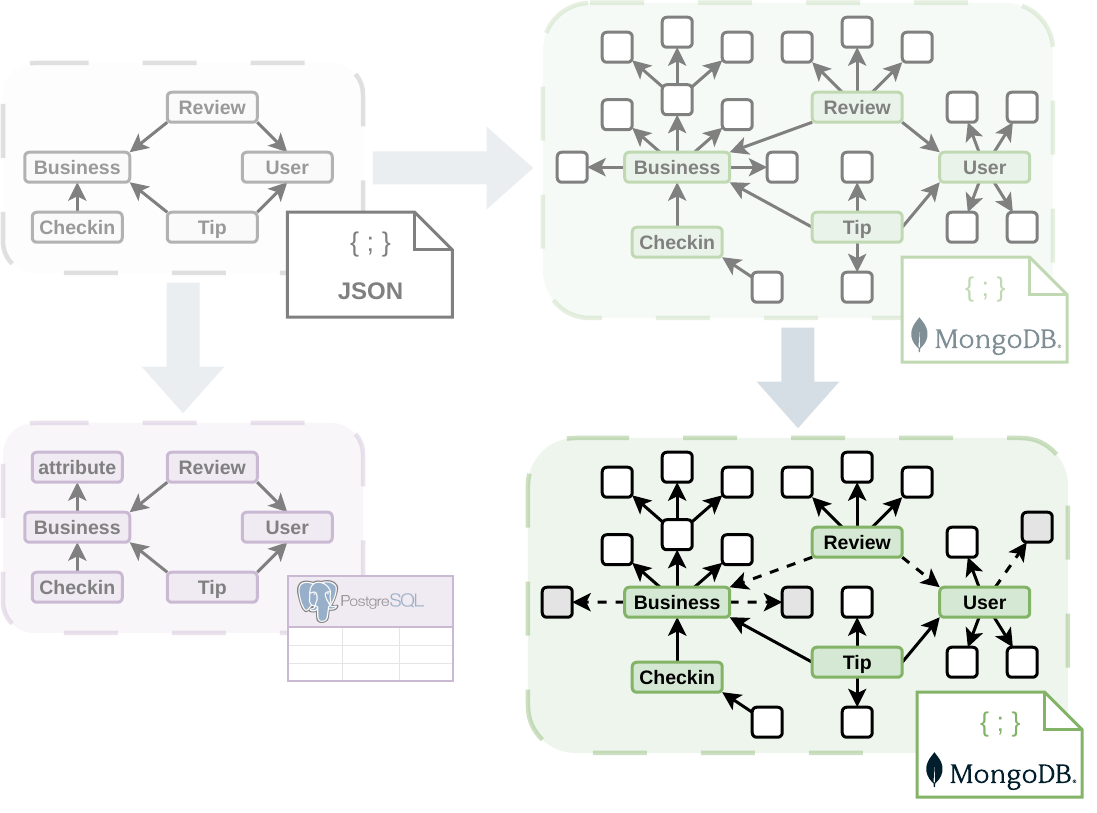}
    \caption{Transformations of the Yelp dataset to embedded MongoDB (and PostgreSQL).}
    \label{fig:drawio/yelp-agg}
\end{figure}

Figure~\ref{fig:drawio/yelp-agg} illustrates the embedding transformation. The schema category remains unchanged; the change is applied at the mapping level. Concretely, we embed business attributes (\textit{state}, \textit{city}, \textit{name}) and user \textit{name} into the \textit{Review} mapping. Queries \textbf{Q2}, \textbf{Q5}, and \textbf{Q6} are modified accordingly.

\begin{figure}
    \centering
    \includegraphics[width=0.9\textwidth]{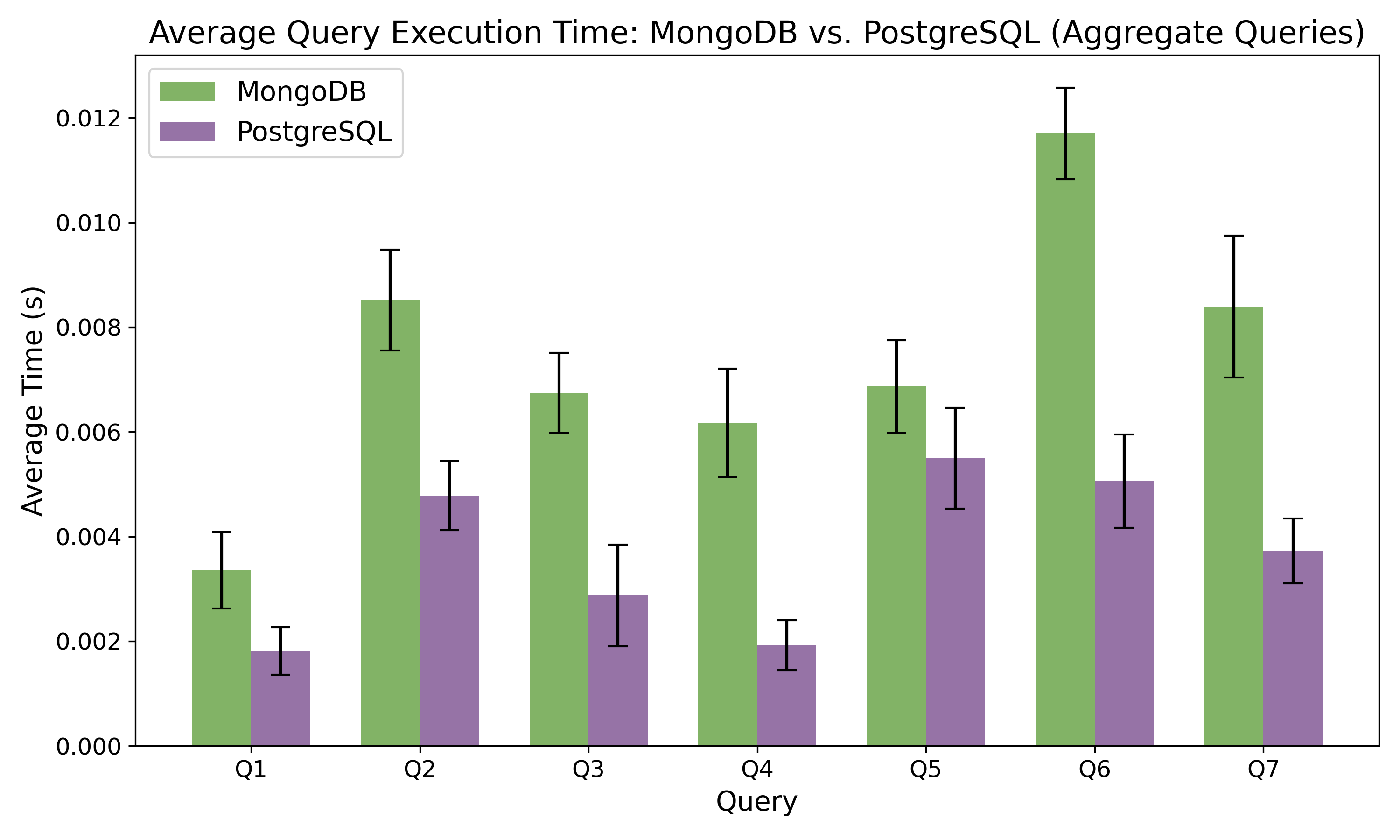}
    \caption{Average execution times of the queries from Table~\ref{tab:yelp-queries} on the Yelp dataset, comparing Embedded MongoDB and PostgreSQL. Error bars represent one standard deviation over the last 50 runs per query.}
    \label{fig:plot/yelp-agg}
\end{figure}

Figure~\ref{fig:plot/yelp-agg} shows that embedding substantially reduces query costs for performance-critical patterns and makes execution times more comparable across representations. This scenario demonstrates that \emph{TransforMMer} can generate controlled benchmark variants reflecting an evolution step driven by workload considerations.

\paragraph{\textbf{Scenario C:} System Evolution through Data Integration and Enrichment}
\label{sec:mm-input}
This scenario reflects the evolution of the system through the integration of additional data sources. We demonstrate multi-model input integration by enriching Yelp business listings with US city information. This reflects a common information systems scenario in which additional data sources are integrated over time.

\begin{figure}
    \centering
    \includegraphics[width=\textwidth]{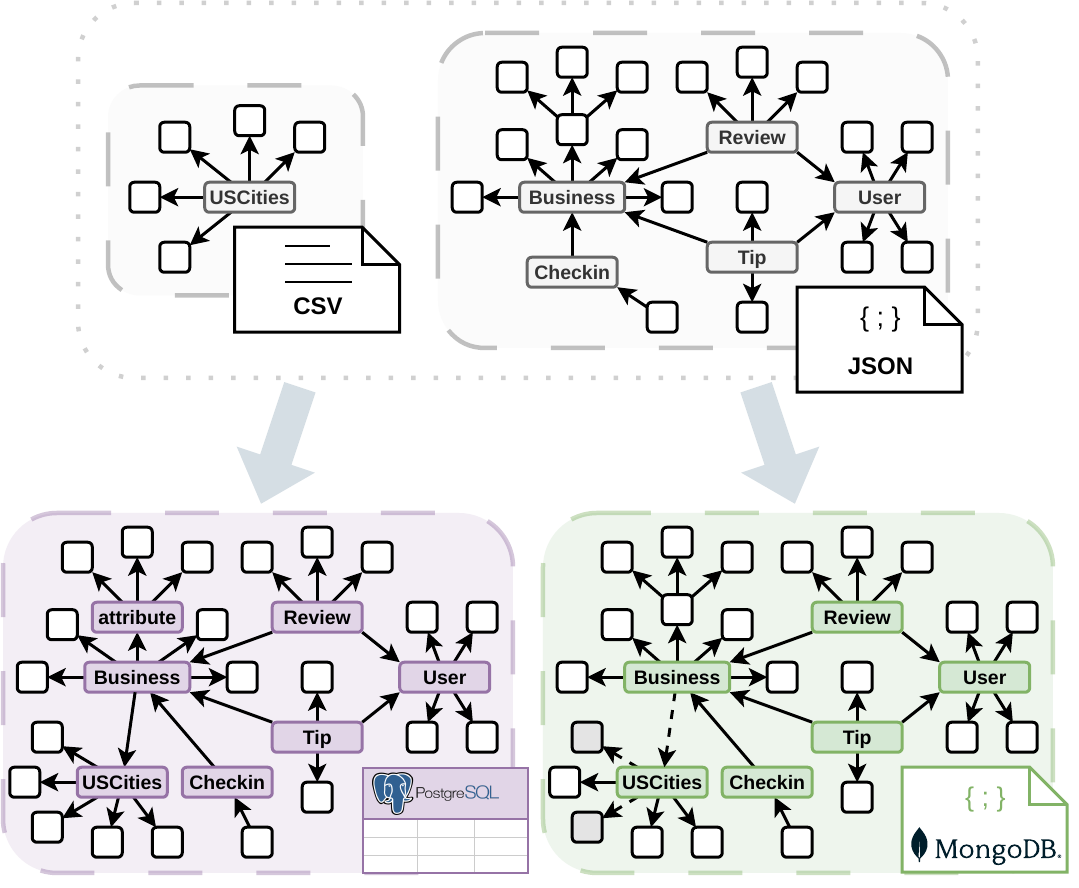}
    \caption{Transformations of the Yelp dataset and US Cities Data to embedded MongoDB (and PostgreSQL).}
    \label{fig:drawio/yelp-uscities}
\end{figure}

Figure~\ref{fig:drawio/yelp-uscities} shows the enrichment and the resulting transformations. We embed selected \textit{USCities} fields into \textit{Business} (\textit{population}, \textit{density}, \textit{timezone}). Table~\ref{tab:yelp-uscities-queries} lists the query set.

\begin{table}[h]
\small
    \caption{Experimental Queries for Yelp dataset with US Cities data}
    \label{tab:yelp-uscities-queries}
    \centering
    \begin{tabular}{@{}cp{6.3cm}p{2.4cm}p{2.6cm}@{}}
      \toprule
      \textbf{Query} & \textbf{Query Description} & \textbf{Type} & \textbf{Note} \\
      \midrule
      \textbf{Q8}  & Find businesses located in cities with a population over one million. & Filter + Join & Enrichment using population \\
      \textbf{Q9}  & Find cities with more than 100 businesses and a population over 500{,}000. & Aggregation + Join & Count + filter with city data \\
      \textbf{Q10} & List all businesses in cities within the Eastern timezone. & Filter + Join & Join with timezone filter \\
      \textbf{Q11} & Count the number of businesses per timezone. & Group + Count & Timezone aggregation \\
      \textbf{Q12} & Top 10 most densely populated cities sorted by average business rating. & Aggregation + Sort + Join & Ranking with enrichment \\
      \bottomrule
    \end{tabular}
\end{table}

\begin{figure}
    \centering
    \includegraphics[width=0.9\textwidth]{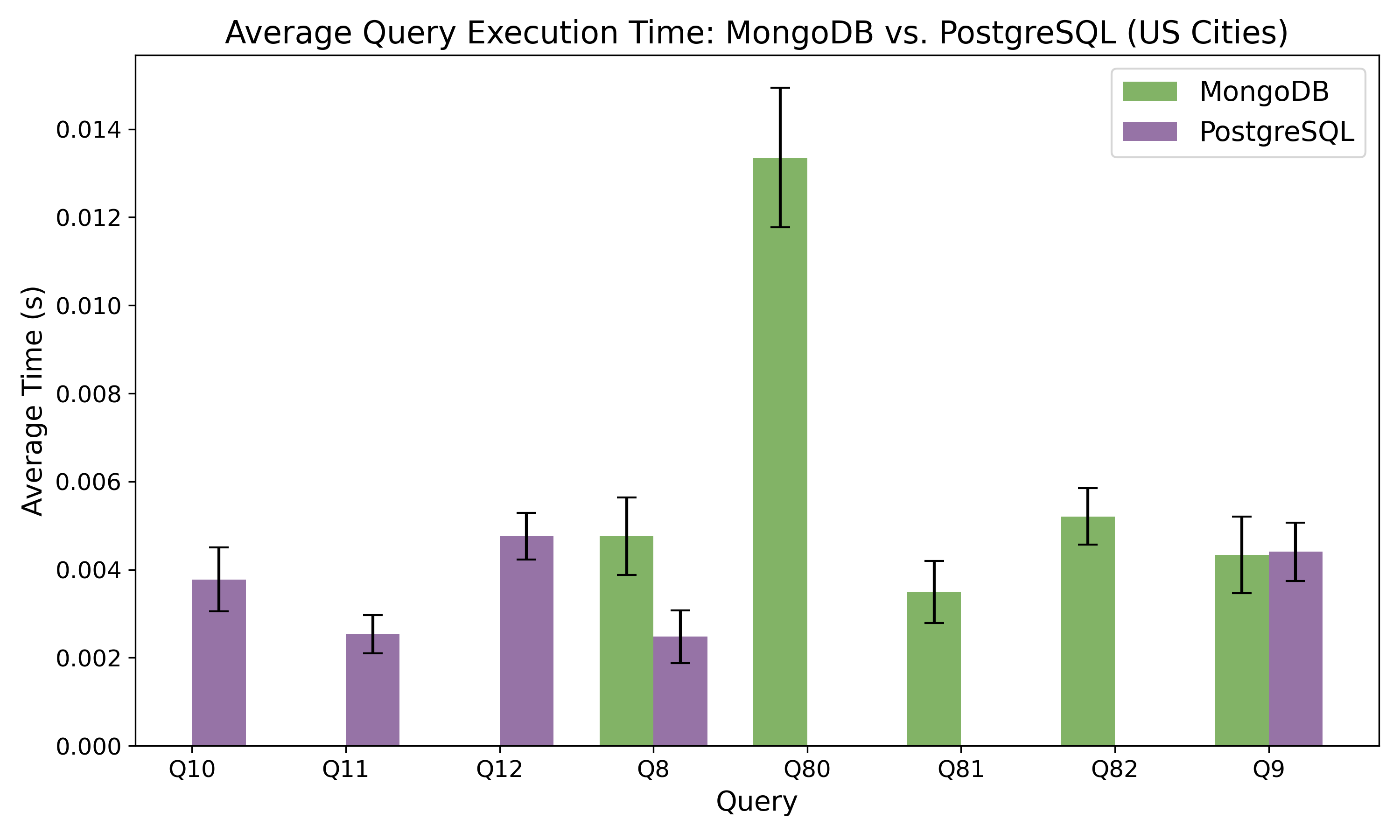}
    \caption{Average execution times of the queries from Table~\ref{tab:yelp-uscities-queries} on the Yelp dataset and US Cities data, comparing Embedded MongoDB and PostgreSQL. Error bars represent one standard deviation over the last 50 runs per query.}
    \label{fig:plot/yelp-uscities}
\end{figure}

Figure~\ref{fig:plot/yelp-uscities} shows that MongoDB tends to be faster on simple filter-heavy queries such as \textbf{Q8} and \textbf{Q10}, likely due to embedding that reduces join-like processing at query time. PostgreSQL is more stable for aggregation-heavy queries such as \textbf{Q9} and \textbf{Q12}, where optimizer and aggregation strategies are more effective. This scenario illustrates how enrichment changes the workload and how performance trade-offs depend on representation decisions.

\subsubsection{BibleData: Relationship-Centric System Representations}

BibleData provides a rich network of relationships, which makes it suitable for demonstrating graph transformations. Table~\ref{tab:bible-queries} lists the query set.

\begin{table}
\small
    \caption{Experimental Queries for BibleData}
    \label{tab:bible-queries}
    \centering
    \begin{tabular}{@{}cp{6.3cm}p{2.4cm}p{2.6cm}@{}}
      \toprule
      \textbf{Query} & \textbf{Query Description} & \textbf{Type} & \textbf{Note} \\
      \midrule
      \textbf{Q13} & Persons mentioned in references describing events in Jerusalem. & Filter + Join & Pattern navigation + filtering \\
      \textbf{Q14} & For each person, count how many unique labels describe them. & Aggregation & Counting and grouping \\
      \textbf{Q15} & Persons connected via a relationship of type ``father''. & Pattern + Filter & Direct traversal \\
      \textbf{Q16} & Persons mentioned together in the same reference. & Join + Group & Co-occurrence detection \\
      \textbf{Q17} & References where a person is in an event and described with a label. & Multi-relation Join & Multi-edge matching \\
      \textbf{Q18} & For each label type, count how many persons it describes. & Group + Count & Aggregation across labels \\
      \textbf{Q19} & Average age of persons at the time of events described in Genesis. & Join + Aggregation & Join-heavy aggregation \\
      \textbf{Q20} & Label trail (Label $\rightarrow$ Verse $\rightarrow$ Person) for King David. & Path Traversal & Deep traversal \\
      \bottomrule
    \end{tabular}
\end{table}

\begin{figure}
    \centering
    \includegraphics[width=\textwidth]{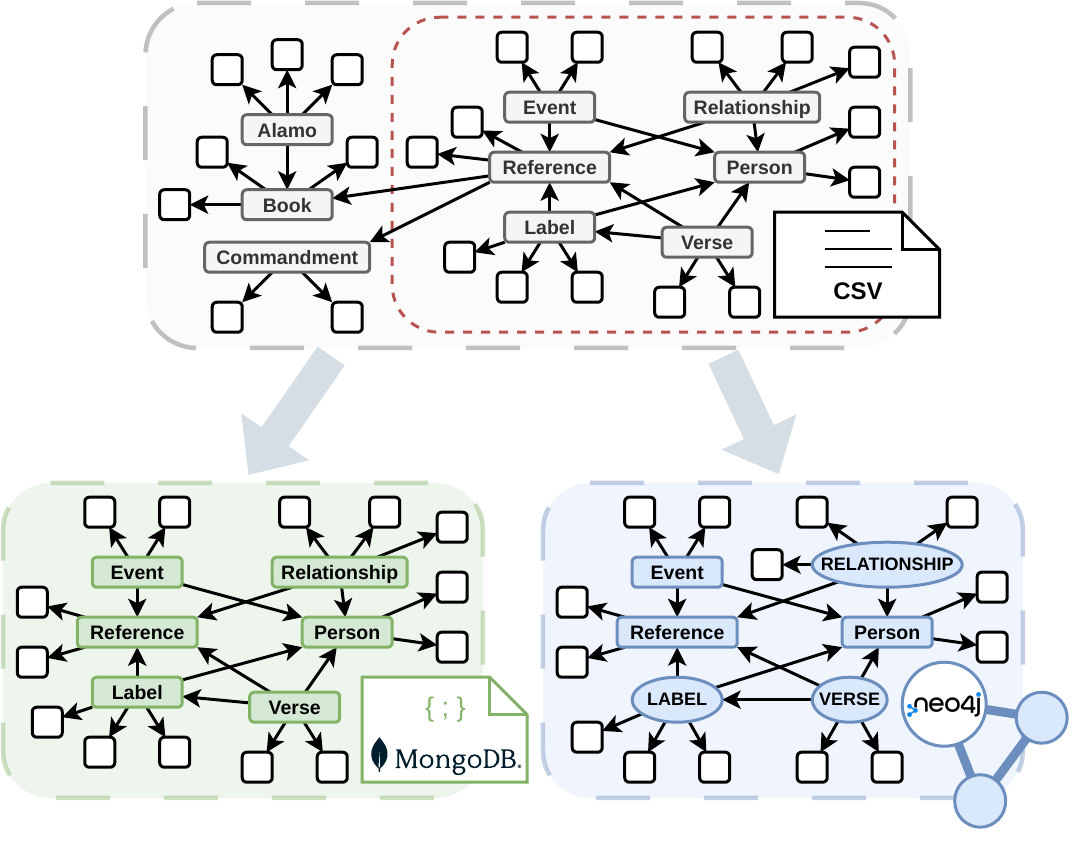}
    \caption{Transformations of the BibleData dataset to MongoDB and Neo4j.}
    \label{fig:drawio/bibledata-split}
\end{figure}

\paragraph{\textbf{Scenario D:} Choosing Between Document-Oriented and Graph-Oriented System Architectures}
\label{sec:mongo-vs-neo4j}
This scenario represents an architectural choice for relationship-centric information systems. To demonstrate a graph transformation, we convert a portion of the dataset to Neo4j and compare it with MongoDB. MongoDB is used for its flexible document schema, which is helpful given the dataset's missing values. We transform only a subset because one part lacks meaningful relationships; this is indicated by the red-highlighted section in Figure~\ref{fig:drawio/bibledata-split}.

Figure~\ref{fig:drawio/bibledata-split} shows that the number of kinds remains the same after both transformations, while Neo4j allows some schema objects to be modeled as relationships (shown as capitalized labels). The \textit{label} plays a dual role depending on the context.

\begin{figure}
    \centering
    \includegraphics[width=0.9\textwidth]{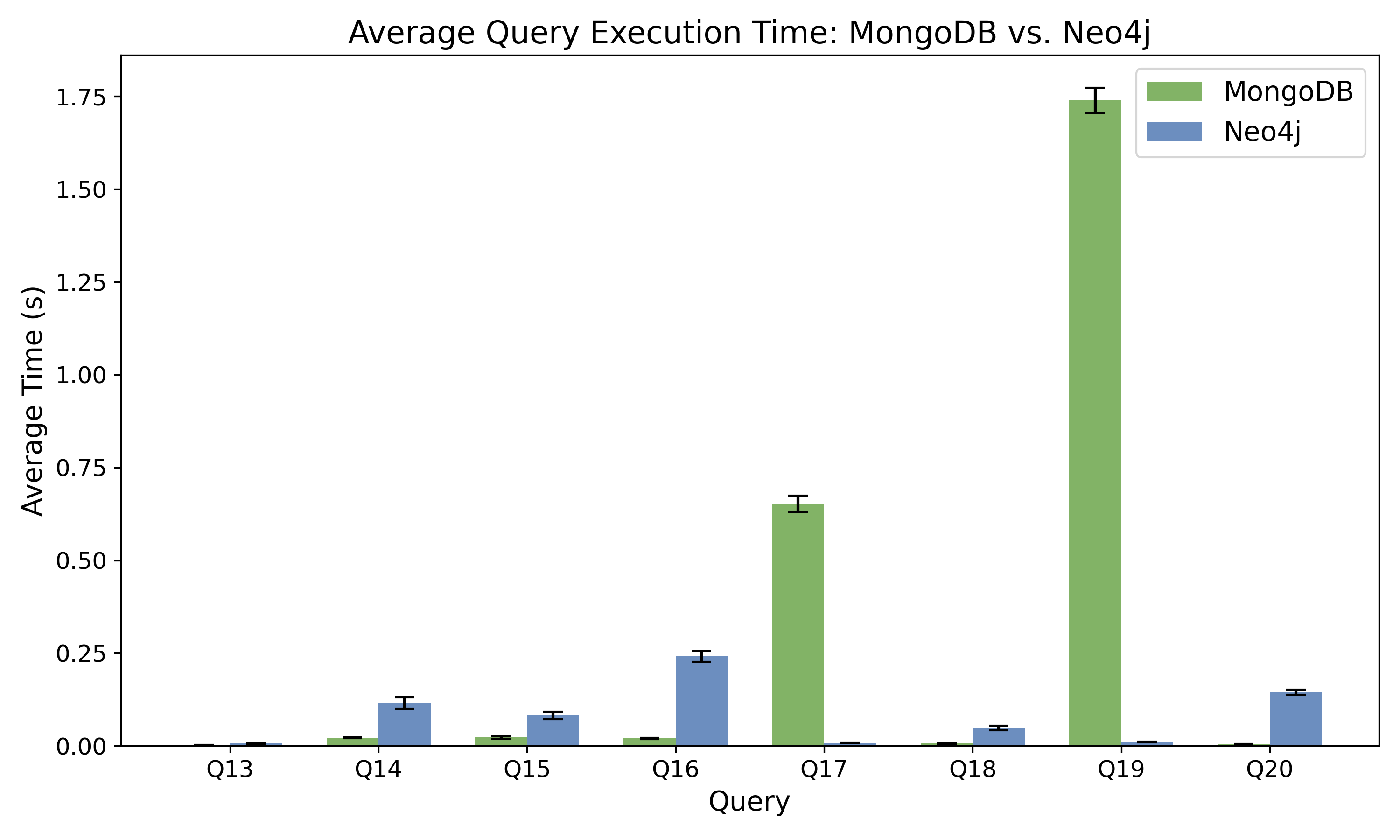}
    \caption{Average execution times of the queries from Table~\ref{tab:bible-queries} on the Bible dataset, comparing MongoDB and Neo4j. Error bars represent one standard deviation over the last 50 runs per query.}
    \label{fig:plot/bibledata-base}
\end{figure}

\begin{figure}
    \centering
    \includegraphics[width=0.9\textwidth]{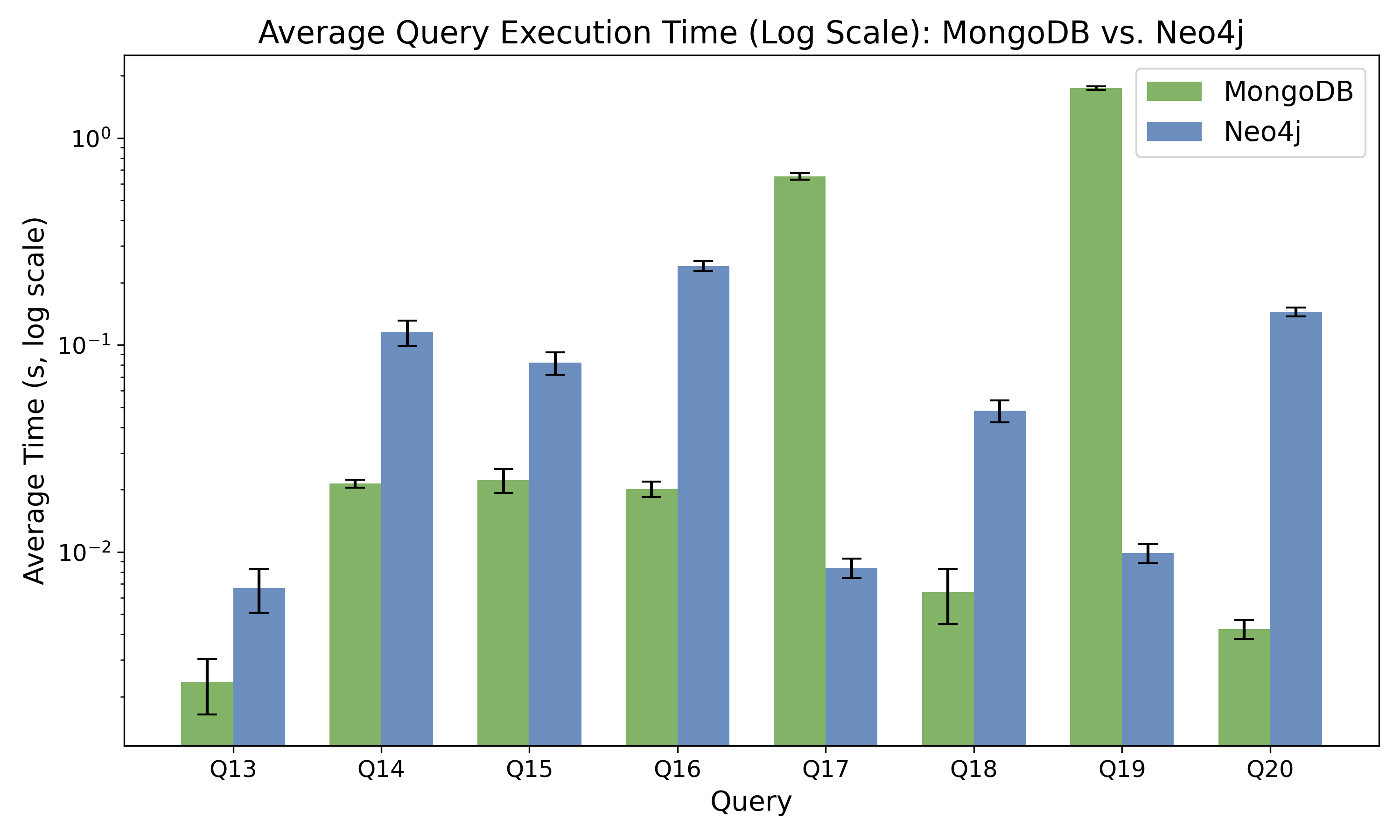}
    \caption{Average execution times of the queries from Table~\ref{tab:bible-queries} on the Bible dataset, comparing MongoDB and Neo4j. Error bars represent one standard deviation over the last 50 runs per query. Time is shown on a logarithmic scale.}
    \label{fig:plot/bibledata-base-log}
\end{figure}

Figures~\ref{fig:plot/bibledata-base} and~\ref{fig:plot/bibledata-base-log} show the measured query costs. MongoDB performs better on simpler queries (Q13--Q17), while Neo4j is more efficient for relationship-intensive queries (Q18--Q20) involving deeper traversal.
\old{This scenario demonstrates that the choice of representation (document vs graph) influences which workloads become efficient and motivates evolution-like redesign decisions.}\new{This scenario demonstrates that the choice of representation (document vs graph) is inherently workload-dependent: MongoDB is more efficient for simpler lookup and aggregation patterns, while Neo4j excels at relationship-intensive traversal workloads. A shift in the system's access patterns---for example, a growing demand for multi-hop graph traversal queries---would therefore justify migrating the relevant portion toward a graph representation, a transition that the framework can generate from the same conceptual schema without modifying it.}

\paragraph{\textbf{Scenario E:} Polyglot Persistence and Hybrid Information System Architecture}
\label{sec:partitions}
Finally, this scenario emulates a polyglot persistence architecture combining multiple specialized data technologies. We emulate a hybrid evolution step by storing different parts of the dataset in separate systems to exploit their respective strengths. We store one part in MongoDB and another in Neo4j. This allows inclusion of the entire dataset, including the previously excluded part, which can be handled efficiently in MongoDB due to its flatter structure.
\new{This partitioned design can also be interpreted as a workload-driven adaptation: the parts of the dataset subject to traversal-intensive access patterns are assigned to Neo4j, while those accessed primarily through flat lookups reside in MongoDB. Both portions are derived from the same conceptual schema, enabling systematic evaluation of such hybrid workload assignments.}

\begin{figure}
    \centering
    \includegraphics[width=\textwidth]{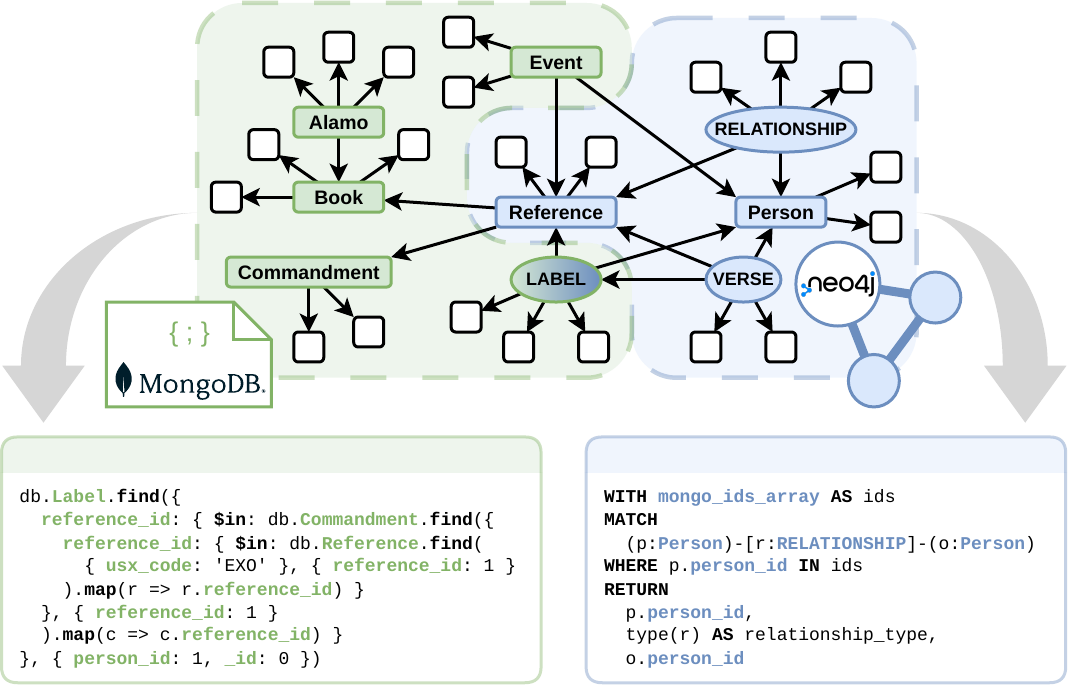}
    \caption{Partitioned transformation with queries on the BibleData dataset in MongoDB and Neo4j.}
    \label{fig:drawio/bibledata-hybrid}
\end{figure}

Figure~\ref{fig:drawio/bibledata-hybrid} illustrates the partitioned transformation and cross-model queries. The green areas represent data stored in MongoDB and the blue areas represent data stored in Neo4j. The intentional overlap in \textit{Label} reflects a redundant but practical design choice to support cross-model access patterns.

\subsection{Summary of Findings and Framework Capabilities}
\label{sec:exp-summary}

Across the scenarios, the experiments demonstrate the following:
\begin{itemize}
    \item Comparable benchmark datasets can be generated across heterogeneous target systems from a shared conceptual basis (supporting the benchmark engineering objective).
    \item Representation choices and evolution-like redesign steps (embedding, enrichment, partitioning) lead to measurable differences in observed query costs and reveal model-specific trade-offs.
    \item The framework enables controlled exploration of structural variants while maintaining comparability of the underlying data and analytical intent.
\end{itemize}

\paragraph{Transformational Capabilities}

Table~\ref{tab:transforms} summarizes the transformation types demonstrated in the scenarios above. Although presented as individual operations, these transformations can be combined to construct more complex benchmark configurations.

\begin{table}[h]
    \caption{Transformational Capabilities of the Proposed Framework}
    \label{tab:transforms}
    \centering
    \begin{tabular}{@{}p{3cm}p{1.5cm}p{1.5cm}p{3cm}p{3cm}@{}}
      \toprule
      \textbf{Transformation} & \textbf{Input Model} & \textbf{Output Model} & \textbf{Key Operations} & \textbf{Experiments} \\
      \midrule
      \textbf{Cross-model} & Single & Single & Converts between data models & \ref{sec:mongo-vs-postgres}, \ref{sec:mongo-vs-neo4j} \\
      \textbf{Multi-Model Generation} & Single & Multi & Creates multiple views from a single source model & \ref{sec:partitions} \\
      \textbf{Aggregation / Embedding} & Single or Multi & Single or Multi & Introduces redundancy to reduce query-time joins & \ref{sec:mongo-embedded-vs-postgres}, \ref{sec:mm-input} \\
      \textbf{Cross multi-model} & Multi & Multi & Unifies sources; reorganizes data across models & \ref{sec:mm-input} \\
      \bottomrule
    \end{tabular}
\end{table}

\subsection{Answering the Research Questions}
\label{sec:rq_answers}

The experimental findings allow us to answer the research questions formulated in Section~\ref{sec:intro}. Rather than interpreting the results as isolated performance comparisons, we analyze them as evidence on how architectural and representation-level decisions influence behavior of heterogeneous information systems.

\paragraph{RQ1: Can multi-model benchmark datasets be systematically generated from heterogeneous real-world data sources using a unified transformation workflow?}

The experiments demonstrate that heterogeneous datasets provided in different formats (e.g., JSON and CSV) can be transformed into multiple structurally distinct \old{yet semantically equivalent representations using a unified transformation workflow.}\new{representations aligned through the same canonical schema and mapping specifications.} Across all scenarios, the framework enabled reproducible generation of relational, document, graph, and hybrid multi-model datasets derived from a shared conceptual representation.

These findings confirm that systematic multi-model benchmark construction from real-world data is feasible and reproducible. More importantly from an Information Systems perspective, they show that representation alternatives can be engineered in a controlled and comparable manner, providing a structured basis for analyzing architectural design choices.

\paragraph{RQ2: How do schema-level transformations and mapping strategies influence observed performance characteristics across heterogeneous systems?}

The results indicate that the representation and mapping strategies significantly influence the observed query behavior. Transformations such as embedding, enrichment, and partitioning altered execution characteristics even when the underlying conceptual data and analytical intent remained identical.

In several scenarios, the same workload produced substantially different performance profiles depending on the structural representation. These findings confirm that schema-level transformation strategies are not merely implementation details but key architectural factors shaping system behavior. From an Information Systems viewpoint, representation design and workload alignment emerge as central elements in architectural decision-making.

\paragraph{RQ3: Can evolution-aware benchmark generation reveal system-level trade-offs that remain hidden in static benchmark settings?}

The experiments show that evolution-like redesign steps---such as attribute embedding, integration of additional data sources, and hybrid partitioning across systems---expose trade-offs that remain invisible in static single-schema evaluations. By generating controlled structural variants derived from the same conceptual basis, the framework makes explicit how changes in representation interact with workload characteristics and system capabilities.

These results demonstrate that evolution-aware benchmark construction enables deeper insight into architectural trade-offs than traditional static benchmarking approaches. It supports systematic exploration of design alternatives under conditions that more closely resemble real-world system evolution.

Overall, the findings suggest that treating benchmark construction as a transformation-driven engineering process enables systematic and reproducible evaluation of heterogeneous information systems under evolving structural conditions. Rather than ranking database technologies, the experiments illustrate how architectural and representation decisions shape system behavior across the life cycle of an information system.

\paragraph{System-Level Interpretation}

From an Information Systems perspective, the framework can be interpreted in multiple complementary ways. First, it serves as a \textbf{benchmark engineering infrastructure} that enables consistent preparation of comparable datasets across heterogeneous platforms. Second, it serves as a \textbf{data integration and transformation layer} that supports cross-model restructuring, enrichment, and partitioning---operations commonly required in modern data engineering and ETL workflows.

More broadly, the framework positions benchmark generation as an analytical instrument for evidence-based architectural evolution. By making structural transformations explicit and reproducible, it supports systematic reasoning about trade-offs in heterogeneous data architectures.

\subsection{Threats to Validity}
\label{sec:threats}

As with any experimental study of complex information systems, the results presented must be interpreted in light of several limitations related to the experimental design, implementation choices, and scope of evaluation. The goal of this section is not to question the observed effects, but to clarify the conditions under which they hold and to outline factors that may influence their generalizability. 

Because the study focuses on representation-aware and evolution-oriented analysis rather than absolute system benchmarking, the primary risks concern the comparability of scenarios, the interpretation of measured behavior, and the transferability of findings to other architectural contexts. The following paragraphs discuss these aspects in terms of internal, construct, external, and conclusion validity.

\paragraph{Internal validity}
To mitigate transient measurement effects, the database containers were restarted before each measurement set, queries were executed repeatedly, and an initial warm-up phase was excluded from the analysis. Nevertheless, Docker scheduling variability and background operating system activity may introduce noise. Since our interpretation focuses on relative trends across structurally distinct variants rather than absolute performance values, the conclusions are less sensitive to minor fluctuations.

\paragraph{Construct validity}
Queries were manually translated into system-specific languages, which can influence execution behavior through formulation and indexing choices. We mitigate this risk by maintaining consistent analytical intent across representations and by ensuring that all systems operate on datasets generated from the same conceptual basis. The purpose is not to demonstrate optimal tuning for each system, but to observe how changes in representation influence system behavior under comparable conditions.

\paragraph{External validity}
The experiments involve three database systems and selected real-world datasets. While numerical results may vary across platforms or larger-scale deployments, the observed patterns---such as the impact of embedding, enrichment, and partitioning---reflect general architectural trade-offs encountered in heterogeneous information systems. Therefore, the conclusions primarily concern methodological implications rather than performance claims about specific technologies.

\paragraph{Conclusion validity}
Execution times were averaged over multiple runs and variability is reported using standard deviation error bars. The interpretation emphasizes consistent behavioral trends across scenarios rather than marginal differences between systems.

\subsection{DaRe}
\label{sec:dare}

To support transparency, reuse, and longitudinal evaluation of architectural decisions, we maintain a repository called \textbf{DaRe} (Dataset Repository)\footnote{\url{https://dare.mmcatdb.com/}}. DaRe stores original datasets, unified schema representations, transformation configurations, mappings, and generated benchmark variants corresponding to the scenarios presented in this paper.

By preserving transformation artifacts and version metadata alongside generated datasets, DaRe enables reproducible reconstruction of alternative structural representations. This supports a systematic comparison of architectural and representation-level design decisions in evolving heterogeneous information systems.

In this sense, DaRe complements the proposed framework by providing an infrastructure for repeatable evaluation and iterative system evolution, enabling both researchers and practitioners to analyze and extend benchmark scenarios under controlled and documented conditions.

\begin{figure}
    \centering
    {\setlength{\fboxsep}{0pt}\fbox{\includegraphics[width=\textwidth,keepaspectratio]{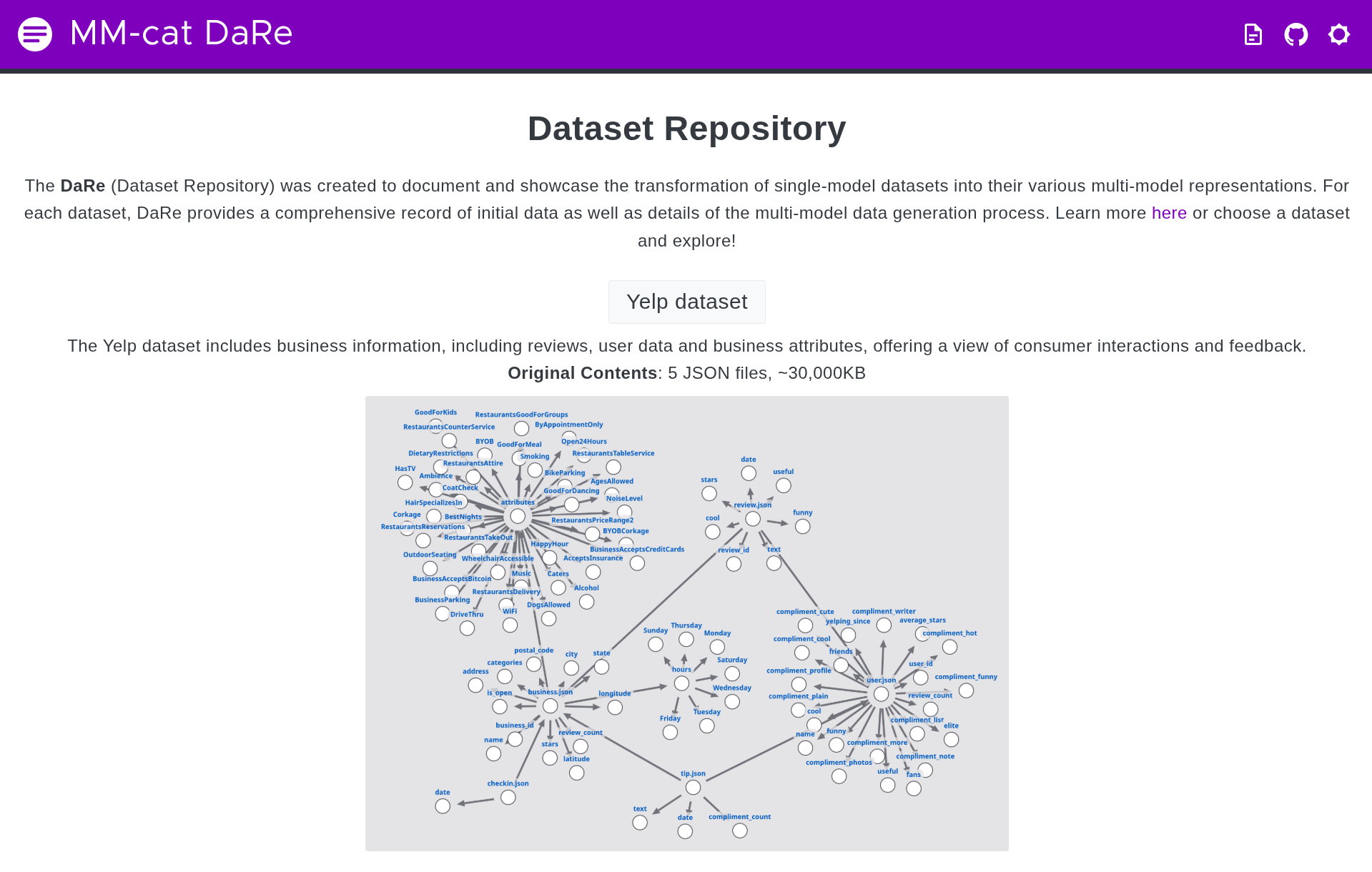}}}
    \caption{DaRe repository supporting reproducible benchmark engineering and evolution-aware evaluation of heterogeneous information systems.}
    \label{fig:dare}
\end{figure}

%% file: chapters/chap06.tex
\section{Discussion and Implications for Information Systems}
\label{sec:discussion}

The results of this study extend beyond database performance comparison and have broader implications for the design and evolution of heterogeneous information systems. Contemporary information systems rarely operate within a single data model or a static architectural configuration. Instead, they evolve incrementally in response to workload changes, integration demands, scalability requirements, and organizational constraints. Structural redesign---such as embedding, normalization changes, enrichment with external data sources, or partitioning across multiple storage technologies---is therefore a recurring architectural activity rather than an exceptional event.

\subsection{Benchmarking as an Architectural Design Instrument}

Traditional benchmarking approaches evaluate database systems under fixed schemas and predefined workloads. While useful for technology comparison, such static benchmarks provide limited support for architectural decision-making in evolving environments. The presented framework repositions benchmarking as an explicit architectural design-support instrument.

By enabling controlled generation of structurally distinct yet semantically \old{equivalent}\new{aligned} dataset variants, the approach allows system architects to explore alternative representation strategies before committing to long-term architectural decisions. Embedding strategies, hybrid partitioning, or cross-model reorganizations can be evaluated systematically rather than introduced through ad hoc experimentation in production systems.

From this perspective, evolution-aware benchmarking becomes part of the information systems design cycle. It supports iterative exploration of structural alternatives, makes performance implications observable, and documents transformation decisions in a reproducible manner.

\subsection{Representation-Aware Trade-offs in Heterogeneous Systems}

The experiments demonstrate that the observed system behavior is not determined solely by database technology. Instead, it emerges from the interaction between representation design, workload characteristics, and architectural context. Join-intensive analytical patterns behave differently under normalized relational schemas, embedded document structures, and graph-based representations. Likewise, hybrid layouts combining multiple systems may exploit complementary strengths.

For information systems architects, this implies that the selection of persistence technology cannot be separated from modeling decisions. Architectural trade-offs must consider the representation strategy, anticipated workload evolution, integration patterns, and long-term maintainability simultaneously. The proposed framework makes such trade-offs explicit by treating representation-level alternatives as configurable, reproducible transformations rather than as implicit implementation choices.

\new{
\subsection{Workload Evolution as a Driver of Structural Adaptation}

Schema changes are a well-recognized driver of structural evolution in heterogeneous information systems. However, changes in application workloads and access patterns represent an equally significant---and in practice often more frequent---source of structural dynamics. New query patterns, shifting analytical demands, or the integration of new data consumers can require restructuring the logical representation of data even when the underlying conceptual schema remains unchanged.

The proposed framework explicitly accommodates this class of adaptation. Because mappings between the schema category and the logical representations are treated as first-class, independently modifiable artifacts, alternative logical representations can be generated by modifying mappings alone, without altering the conceptual schema. This decoupling of conceptual stability from representational flexibility is a key design property of the framework.

More broadly, for any significant workload change the framework enables systematic exploration of which logical representation best accommodates the new access patterns, evaluation of the associated performance trade-offs, and reproducible generation of the adapted configuration. This extends the framework's applicability beyond schema-evolution-driven benchmarking to the systematic analysis of workload-induced structural adaptation throughout the operational life cycle of heterogeneous information systems.
}

\subsection{Implications for the Information Systems Life Cycle}

A further contribution concerns life cycle integration. In practice, information systems undergo continuous redesign: new data sources are integrated, schemas evolve, and storage strategies are adapted. However, the evaluation of these changes is often informal and difficult to reproduce.

By storing transformation artifacts, mappings, and dataset versions alongside metadata, the framework enables traceable, repeatable evaluation of architectural modifications. This capability aligns benchmarking with iterative system development processes and enables evidence-based architectural evolution across the design, integration, optimization, and reconfiguration phases.

Rather than treating benchmarking as a one-time post-deployment activity, the presented approach integrates it into the broader life cycle of heterogeneous information systems. In this sense, transformation-driven benchmark engineering provides a structured bridge between conceptual data modeling and empirical system evaluation.

\subsection{Positioning Within Information Systems Research}

Within the Information Systems discipline, this work's contribution lies in its methodological shift. Instead of proposing a new database engine or optimization technique, the framework introduces a structured process for analyzing architectural alternatives in evolving multi-model environments. It operationalizes representation-aware experimentation and connects formal modeling foundations with practical system-level evaluation.

The study therefore contributes to research on system design methodologies, architectural trade-off analysis, life-cycle-aware evaluation, and reproducible experimentation in heterogeneous information systems. By formalizing structural transformations and making them experimentally observable, the framework supports systematic reasoning about architectural decisions under evolving conditions.

%% file: chapters/epilog.tex
\section{Conclusion}
\label{sec:concl}

This work positions evolution-aware benchmark construction as a design-support methodology for heterogeneous information systems rather than merely as a performance comparison tool. In modern multi-model environments, architectural decisions regarding data representation, integration, and storage strategies directly influence system behavior, maintainability, and the cost of evolution. However, such decisions are often evaluated informally or retrospectively.

We introduced \emph{TransforMMer}, a transformation-driven framework grounded in a unified structural representation that enables systematic generation of comparable benchmark variants across relational, document, and graph systems. By treating schema transformations and representation changes as explicit, configurable artifacts, the framework supports reproducible exploration of architectural alternatives.

The experimental scenarios demonstrated that evolution-like redesign steps---such as embedding, enrichment, and partitioned storage---produce measurable and explainable differences in observed behavior. Crucially, these differences arise from the interaction between workload characteristics and representation design, rather than from database technology alone.
\new{This highlights a broader applicability of the framework: beyond propagating conceptual schema changes to logical representations and queries, it also supports workload-driven adaptation of logical representations while preserving the underlying schema---a scenario that is arguably more frequent in production systems than schema restructuring itself.} This highlights the importance of representation-aware evaluation in heterogeneous information systems.

Beyond its technical mechanisms, this work's primary contribution is methodological. It reframes benchmarking as an integral component of the information systems life cycle, enabling controlled experimentation with structural alternatives before making architectural commitments. By integrating conceptual modeling, structural transformation, and empirical evaluation into a coherent process, the proposed approach supports evidence-based architectural evolution in complex data ecosystems.

Future work will extend the framework with automated query transformation, unified workload specification, \new{systematic benchmarking of workload-driven structural adaptations,} and broader evaluation across distributed deployments. We envision transformation-driven benchmark engineering as a reusable methodological foundation for systematic architectural analysis and life-cycle-aware evolution of heterogeneous information systems.

%% file: bibliography.bib
@article{Kim2022,
    title        = {{M2Bench}},
    author       = {Bogyeong Kim and Kyoseung Koo and Undraa Enkhbat and Sohyun Kim and Juhun Kim and Bongki Moon},
    year         = 2022,
    month        = {dec},
    journal      = {Proceedings of the {VLDB} Endowment},
    publisher    = {Association for Computing Machinery ({ACM})},
    volume       = 16,
    number       = 4,
    pages        = {747--759}
}

@misc{TPC,
    title        = {TPC Benchmarks Overview},
    author       = {{Transaction Processing Performance Council}},
    howpublished = {\url{https://www.tpc.org/information/benchmarks5.asp}},
    note         = {Accessed: 2026-02-22},
    year         = {2026},
    organization = {Transaction Processing Performance Council},
    url          = {https://www.tpc.org/information/benchmarks5.asp}
}

@inproceedings{LDBC,
    author       = {Erling, Orri and Averbuch, Alex and Larriba-Pey, Josep and Chafi, Hassan and Gubichev, Andrey and Prat, Arnau and Pham, Minh-Duc and Boncz, Peter},
    title        = {The LDBC Social Network Benchmark: Interactive Workload},
    year         = {2015},
    isbn         = {9781450327589},
    publisher    = {Association for Computing Machinery},
    address      = {New York, NY, USA},
    pages        = {619-630},
    numpages     = {12},
    location     = {Melbourne, Victoria, Australia},
    series       = {SIGMOD '15}
}

@misc{Faker,
    title        = {Faker Data Generator},
    author       = {{EaseCloud}},
    year         = {2026},
    howpublished = {\url{https://www.easecloud.io/tools/database/faker-data-generator/}},
    note         = {Accessed: 2026-02-22},
    organization = {EaseCloud},
    url          = {https://www.easecloud.io/tools/database/faker-data-generator/}
}

@misc{SDV,
    author       = {Marujo, Carlos},
    title        = {Synthetic Data Generation with SDV Tutorial},
    year         = {2023},
    howpublished = {\url{https://www.kaggle.com/code/mcarujo/synthetic-data-generation-sdv-tutotial}},
    note         = {Kaggle notebook. Accessed: 2026-02-22},
    organization = {Kaggle},
    url          = {https://www.kaggle.com/code/mcarujo/synthetic-data-generation-sdv-tutotial}
}

@inproceedings{YCSB,
    author       = {Cooper, Brian F. and Silberstein, Adam and Tam, Erwin and Ramakrishnan, Raghu and Sears, Russell},
    title        = {Benchmarking cloud serving systems with YCSB},
    year         = {2010},
    isbn         = {9781450300360},
    publisher    = {Association for Computing Machinery},
    address      = {New York, NY, USA},
    booktitle    = {Proceedings of the 1st ACM Symposium on Cloud Computing},
    pages        = {143-154},
    numpages     = {12},
    location     = {Indianapolis, Indiana, USA},
    series       = {SoCC '10}
}

@article{Zhang2019,
    title        = {{Holistic evaluation in multi-model databases benchmarking}},
    author       = {Chao Zhang and Jiaheng Lu},
    year         = 2019,
    month        = {dec},
    journal      = {Distributed and Parallel Databases},
    publisher    = {Springer Science and Business Media {LLC}},
    volume       = 39,
    number       = 1,
    pages        = {1--33}
}

@article{Koupil2022,
    title        = {{A unified representation and transformation of multi-model data using category theory}},
    author       = {Pavel Koupil and Irena Holubov{\'{a}}},
    year         = 2022,
    month        = {may},
    journal      = {Journal of Big Data},
    publisher    = {Springer Science and Business Media {LLC}},
    volume       = 9,
    number       = 1
}

@article{Koupil2022a,
    title        = {{A universal approach for multi-model schema inference}},
    author       = {Pavel Koupil and Sebasti{\'{a}}n Hricko and Irena Holubov{\'{a}}},
    year         = 2022,
    month        = {aug},
    journal      = {Journal of Big Data},
    publisher    = {Springer Science and Business Media {LLC}},
    volume       = 9,
    number       = 1
}

@inproceedings{Koupil2022c,
    title        = {{MM-evocat: A Tool for Modelling and Evolution Management of Multi-Model Data}},
    author       = {Koupil, Pavel and Bártík, Jáchym and Holubová, Irena},
    year         = 2022,
    month        = oct,
    booktitle    = {Proceedings of the 31st ACM International Conference on Information; Knowledge Management},
    publisher    = {ACM},
    address      = {New York, NY, USA},
    series       = {CIKM ’22}
}

@inproceedings{schmidt2002xmark,
    title        = {{XMark: A Benchmark for XML Data Management}},
    author       = {Albrecht Schmidt and Florian Waas and Martin Kersten and Michael J. Carey and Ioana Manolescu and Ralph Busse},
    year         = 2002,
    booktitle    = {Proceedings of the 28th International Conference on Very Large Data Bases (VLDB)},
    publisher    = {Morgan Kaufmann},
    address      = {Hong Kong, China},
    pages        = {974--985},
    
}

@inproceedings{guo2005lubm,
    title        = {{LUBM: A benchmark for OWL knowledge base systems}},
    author       = {Guo, Yuanbo and Pan, Zhengxiang and Heflin, Jeff},
    year         = 2005,
    booktitle    = {International Semantic Web Conference},
    pages        = {123--138},
    organization = {Springer}
}

@article{bagan2016gmark,
    title        = {{gMark: Schema-driven graph query benchmark generation}},
    author       = {Bagan, Guillaume and Bonifati, Angela and Ciucanu, Radu and Fletcher, George H L and Lemay, Aur{\'e}lien and Prat-P{\'e}rez, Arnau},
    year         = 2016,
    journal      = {IEEE Transactions on Knowledge and Data Engineering},
    publisher    = {IEEE},
    volume       = 29,
    number       = 4,
    pages        = {856--869}
}

@inproceedings{ghazal2013bigbench,
    title        = {{BigBench: Towards an Industry Standard Benchmark for Big Data Analytics}},
    author       = {Ahmad Ghazal and Tilmann Rabl and Minqing Hu and Francois Raab and Meikel Poess and Alain Crolotte and Hans-Arno Jacobsen},
    year         = 2013,
    booktitle    = {Proceedings of the 2013 ACM SIGMOD International Conference on Management of Data},
    publisher    = {ACM},
    address      = {New York, NY, USA},
    pages        = {1197--1208}
}

@inproceedings{lengweiler2023mmsbench,
    title        = {{MMSBench-Net: Scenario-Based Evaluation of Multi-Model Database Systems}},
    author       = {Lengweiler, David and Vogt, Marco and Schuldt, Heiko},
    year         = 2023,
    booktitle    = {Proceedings of the 34th GI-Workshop on Foundations of Databases},
    organization = {University of Basel}
}

@inproceedings{Zaharia2010Spark,
    title        = {{Spark: Cluster Computing with Working Sets}},
    author       = {Matei Zaharia and Mosharaf Chowdhury and Michael J. Franklin and Scott Shenker and Ion Stoica},
    year         = 2010,
    booktitle    = {Proceedings of the 2nd USENIX Conference on Hot Topics in Cloud Computing (HotCloud'10)},
    pages        = 10,
}

@misc{zhang2014yelp,
    title        = {{Yelp Challenge Project Report}},
    author       = {Tingting Zhang and Yi Pan},
    year         = 2014,
    note         = {Project report based on the Yelp dataset challenge},
    institution  = {University of Washington},
    howpublished = {\url{https://www.yelp.com/dataset_challenge/}}
}

@misc{json,
    title        = {{The application/json Media Type for JavaScript Object Notation (JSON)}},
    author       = {Douglas Crockford},
    year         = 2006,
    note         = {RFC 4627},
    howpublished = {\url{https://www.rfc-editor.org/rfc/rfc4627}}
}

@misc{alam2021yelp,
    title        = {{Yelp Dataset Analysis using Scalable Big Data}},
    author       = {Alam, Mohsen and Cevallos, Benjamin and Flores, Oscar and Lunetto, Randall and Yayoshi, Kotaro and Woo, Jongwook},
    year         = 2021,
    url          = {https://arxiv.org/abs/2104.08396},
    note         = {Supported by Oracle Cloud Innovation Accelerator},
    institution  = {California State University Los Angeles}
}

@manual{mongodb,
    title        = {{MongoDB: The Developer Data Platform}},
    author       = {{MongoDB Inc.}},
    year         = 2024,
    url          = {https://www.mongodb.com},
    note         = {Version 7.0}
}

@manual{postgresql,
    title        = {{PostgreSQL: The World's Most Advanced Open Source Relational Database}},
    author       = {{The PostgreSQL Global Development Group}},
    year         = 2024,
    url          = {https://www.postgresql.org},
    note         = {Version 16}
}

@misc{docker,
    title        = {{Docker: Lightweight Linux Containers for Consistent Development and Deployment}},
    author       = {{Docker, Inc.}},
    year         = 2013,
    note         = {Accessed: 2025-05-09},
    howpublished = {\url{https://www.docker.com}}
}

@article{makris2020mongodb,
    title        = {{MongoDB vs PostgreSQL: A comparative study on performance aspects}},
    author       = {Antonios Makris and Konstantinos Tserpes and Giannis Spiliopoulos and Dimosthenis Anagnostopoulos},
    year         = 2020,
    journal      = {GeoInformatica},
    volume       = 24,
    number       = 2,
    pages        = {243--268},
    url          = {https://doi.org/10.1007/s10707-020-00407-w}
}

@inproceedings{bartik2025transformmer,
    title        = {{TransforMMer: A Universal Multi-Model Data Generator}},
    author       = {Jáchym Bártík and Alžběta Šrůtková and Irena Holubová},
    year         = 2025,
    pages        = {1150-1153},
    booktitle    = {Proceedings of the 28th International Conference on Extending Database Technology (EDBT)},
    publisher    = {OpenProceedings.org},
    address      = {Barcelona, Spain}
    
}

@inproceedings{holubova2025reshaping,
    title        = {{Reshaping Reality: Creating Multi-Model Data and Queries from Real-World Inputs}},
    author       = {Irena Holubová and Alžběta Šrůtková and Jáchym Bártík},
    booktitle    = {Proceedings of the 20th International Conference on Evaluation of Novel Approaches to Software Engineering, {ENASE} 2025},
    pages        = {174--184},
    publisher    = {{SCITEPRESS}},
    year         = {2025},
    address      = {Porto, Portugal}
}

@article{koupil_universal_2025,
    title        = {{A Universal Approach for Simplified Redundancy-Aware Cross‑Model Querying}},
    author       = {Koupil, Pavel and Crha, Daniel and Holubová, Irena},
    year         = 2025,
    journal      = {Information Systems},
    volume       = 127,
    pages        = 102456,
    url          = {https://doi.org/10.1016/j.is.2024.102456}
}

@inproceedings{felter2015performance,
    title        = {{An updated performance comparison of virtual machines and Linux containers}},
    author       = {Felter, Wes and Ferreira, Alexandre and Rajamony, Ram and Rubio, Juan},
    year         = 2015,
    booktitle    = {2015 IEEE International Symposium on Performance Analysis of Systems and Software (ISPASS)},
    pages        = {171--172},
    organization = {IEEE}
}
